\documentclass[preprint,pre,superscriptaddress,showpacs]{revtex4-1}
\usepackage{graphicx,color}
\usepackage{dcolumn}
\usepackage{amsmath,amssymb}
\usepackage{longtable}
\usepackage{mathtools}
\usepackage{cases}
\usepackage{booktabs}
\usepackage[capitalize]{cleveref}

\graphicspath{./Figure/}
\newcommand{\CHK}[1]{\textcolor{black}{#1}}

\newcommand{\exFig}[1]{\CHK{Fig.~\!{#1}}}
\newcommand{\exEq}[1]{\CHK{Eq.~\!{#1}}}
\newcommand{\exEqs}[1]{\CHK{Eqs.~\!{#1}}}
\newcommand{\exSec}[1]{\CHK{Sec.~\!{#1}}}
\newcommand{\exChap}[1]{\CHK{Chap.~\!{#1}}}
\newcommand{\exApp}[1]{\CHK{Appendix~\!{#1}}}

\newcommand{\myRef}[1]{Ref.~\!\cite{#1}}
\newcommand{\myRefs}[1]{Refs.~\!\cite{#1}}
\newcommand{\myEq}[1]{Eq.~\!(\ref{#1})}
\newcommand{\myEqs}[1]{Eqs.~\!(\ref{#1})}
\newcommand{\myEquation}[1]{Equation~\!(\ref{#1})}
\newcommand{\mySec}[1]{Sec.~\!\ref{#1}}
\newcommand{\mySecs}[1]{Secs.~\!\ref{#1}}
\newcommand{\myFig}[1]{Fig.~\!\ref{#1}}

\newcommand{\myTable}[1]{Table~\!\ref{#1}}
\newcommand{\myAppendix}[1]{Appendix~\!\ref{#1}}
\newcommand{\myav}[1]{\left\langle#1\right\rangle}

\def\myIm{{\rm i}}

\def\mykB{k_{\rm B}}
\def\myTc{T_{\rm C}}
\def\myann{{\cal D}}
\def\mysn{{\rm sn}}
\def\mycn{{\rm cn}}
\def\mydn{{\rm dn}}

\newcommand{\mySn}[1]{\mysn\!\left(#1\right)}
\newcommand{\mysnsn}[3]{k\mySn{\!#1#2}\mySn{\!#1#3}}
\newcommand{\mySnSn}[3]{\left[\mysnsn{#1}{#2}{#3}\right]}
\graphicspath{{./Figure/}}
\newcommand\FIGone{%
 \begin{figure}[tb]
  \begin{center}
   \includegraphics[width=0.5\linewidth]{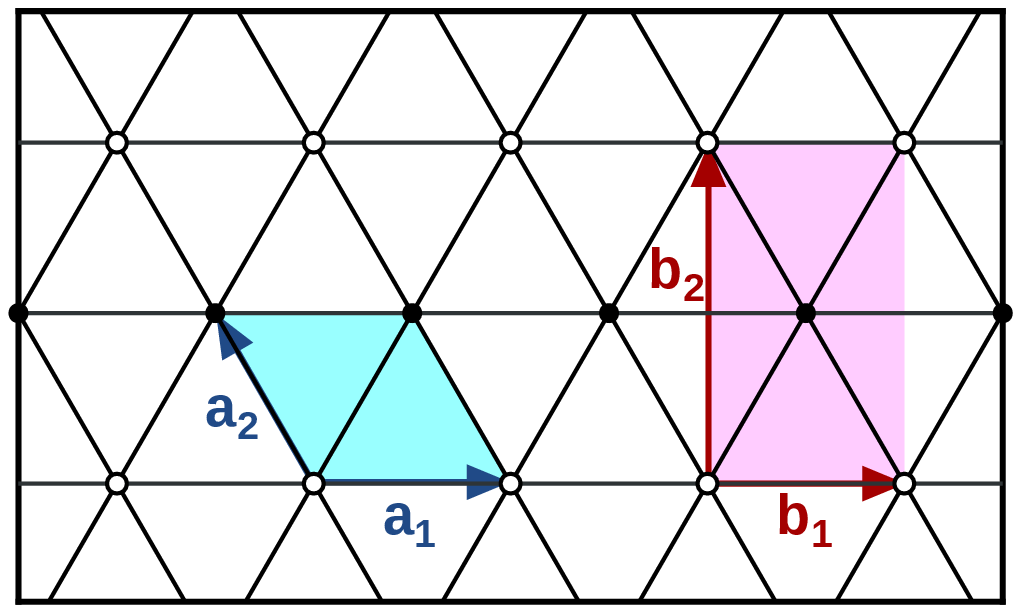}
  \end{center}
  \caption{%
    Blue arrows ${\bf a}_1$ and ${\bf a}_2$ represent the primitive vectors of the triangular lattice, 
    which correspond to ${\cal A}_1(\Theta)$ and ${\cal A}_2(\Theta)$ in \myEq{eq:2.7}, respectively.   
    We divide the triangular lattice into two sublattices, represented by open and closed circles.
    Red arrows ${\bf b}_1$ and ${\bf b}_2$ represent the primitive vectors of a sublattice, 
    which correspond to ${\cal X}(\Theta)$ and ${\cal Y}(\Theta)$ in \myEq{eq:2.1}, respectively. 
  }
  \label{fig:1}
 \end{figure}
 }
 \newcommand\FIGtwo{%
 \begin{figure}[tb]
  \begin{center}
   \includegraphics[width=0.8\linewidth]{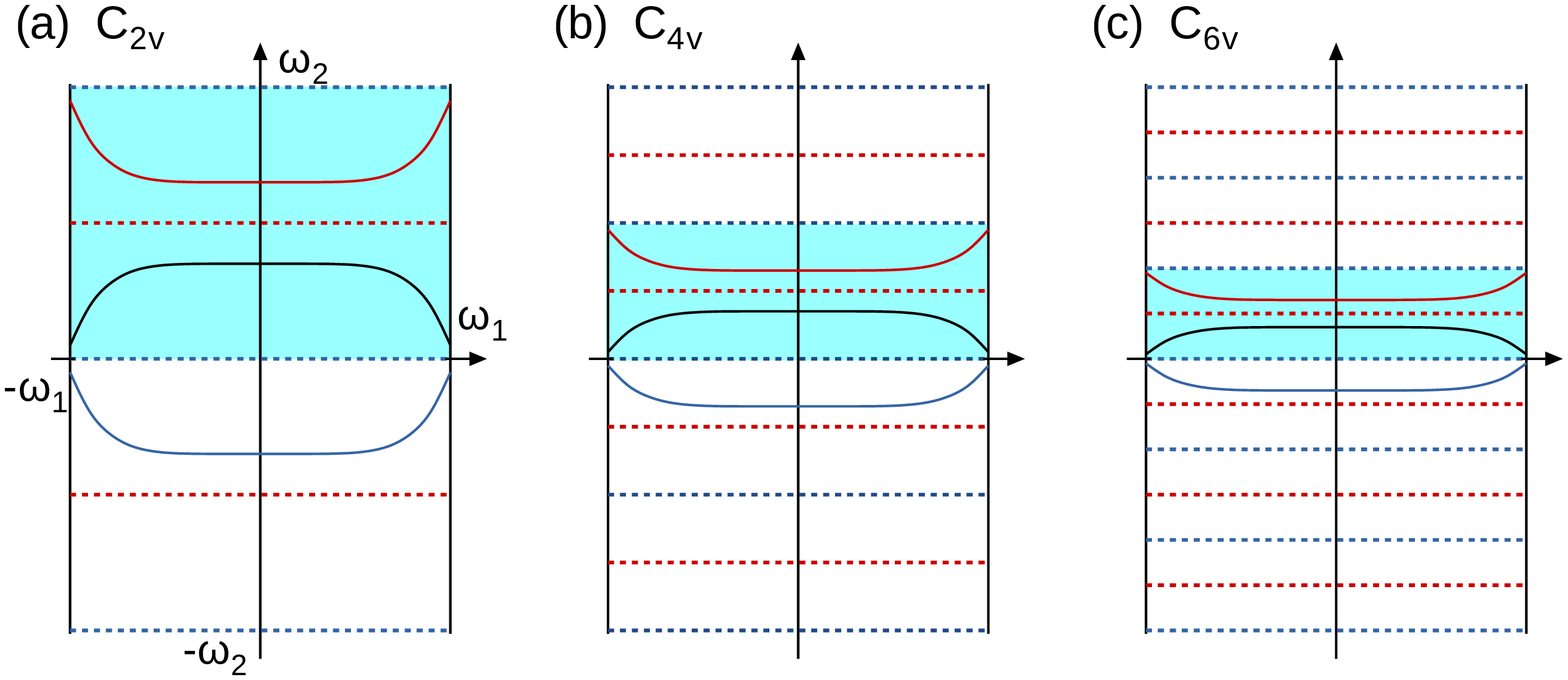}
  \end{center}
  \caption{%
  Periodic rectangles for (a) $C_{2v}$, (b) $C_{4v}$, and (c) $C_{6v}$.
  They are divided into (a) two, (b) four, and (c) six sub-regions because of the rotational symmetries; a cyan area represents a sub-region. 
  Blue and red dotted lines are the reflection axes for the integration paths (see the text).
  For instance, (a) integrals along the paths represented by the black, blue, and red curves yield the same result for \myEq{eq:2.1} if $m$ and $n$ are transformed correspondingly. 
  The same occurs (b) in \exEq{(2.3)} in \myRef{Fujimoto2020} for $C_{4v}$, and (c) in \myEq{eq:2.6} for $C_{6v}$. 
 }
  \label{fig:2}
 \end{figure}
 }
 \newcommand\FIGcolorMap{%
 \begin{figure}[t]
  \begin{center}
   \includegraphics[width=0.98\linewidth]{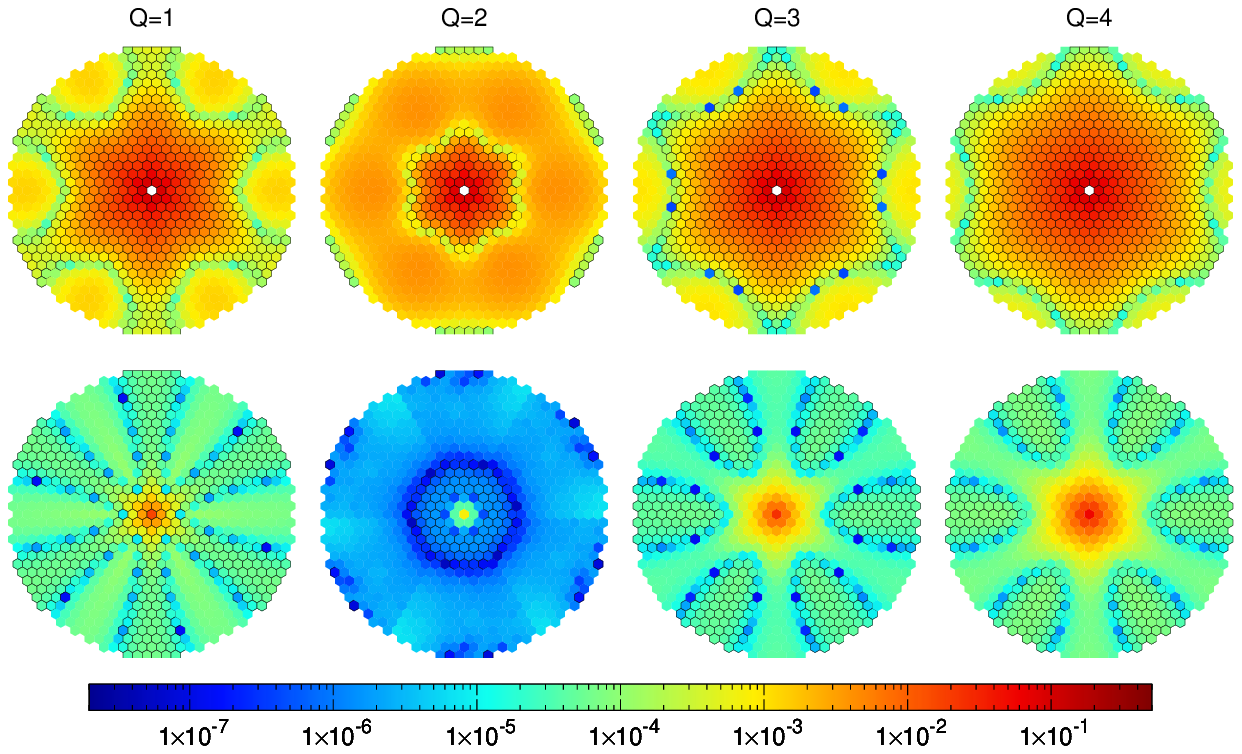}
  \end{center}
  \caption{%
 First and second rows show the color maps of the residual errors ${\cal R}_{\rm OZ}(i,j)$ and ${\cal R}_{\rm tri}(i,j)$, respectively. 
 From left to right, their comparisons are given for $Q=1$, 2, 3, and $4$ at $t=1$, 0.5, 0.3, and 0.2, respectively. 
 Each site $(i,j)$ corresponds to one hexagon whose color represents
 $|{\cal R}_{\rm OZ}|$ or $|{\cal R}_{\rm tri}|$. 
Boundary lines of hexagons indicate that the residual errors are positive.
 }
  \label{fig:colorMap}
 \end{figure}
 }
\newcommand\FIGAone{%
\begin{figure}[t]
\begin{center}
\includegraphics[width=0.6\linewidth]{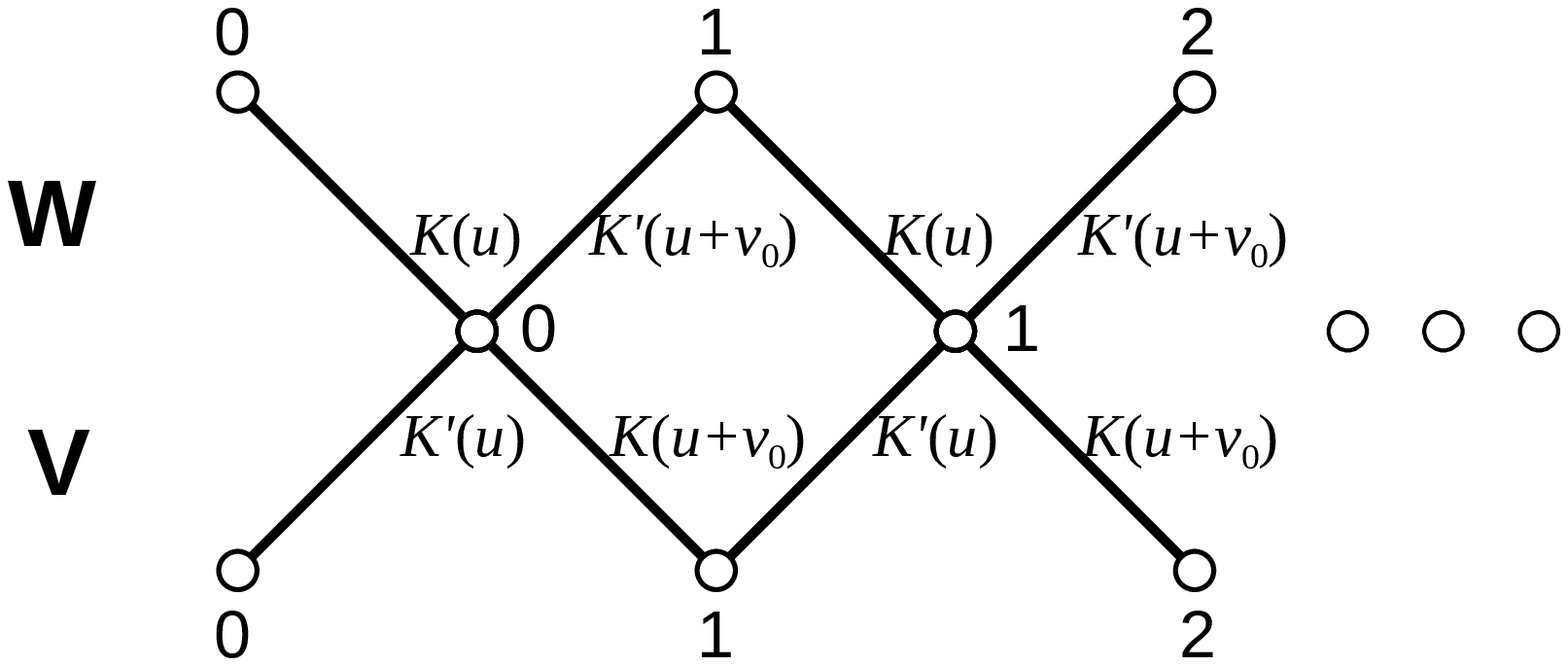}
\end{center}
\caption{%
  Square lattice drawn diagonally. Transfer matrices ${\bf W}$ and ${\bf V}$ connect three successive rows.
  For the parameterization of coupling constants, see the text.
  }
\label{fig:figA1}
 \end{figure}
}
\newcommand\TABLEone{%
 \begin{table*}
  \caption{%
  \label{table:1}
    First column shows the elements in $C_{2v}$: $\pi$-rotation $c_2$, vertical reflection $\sigma_x$, and horizontal reflection $\sigma_y$; second shows the corresponding coordinate transformations; and third shows the induced transformations of the integration path in \myEq{eq:2.1}. 
    Conditions in the fourth column are required for ${\cal X}(\Theta)$ and ${\cal Y}(\Theta)$ to ensure $C_{2v}$ symmetry.
  }
  \begin{ruledtabular}
   \begin{tabular}{cllc}
   $C_{2v}$&
   Coordinate trans.&
   Path shifts&
   Conditions for ${\cal X}(\Theta)$ and ${\cal Y}(\Theta)$
   \\
   \hline
    $c_2$& 
    $(n',m')=(-n,-m)$&
    $\Theta'=\Theta+\omega_2$&
    $\displaystyle{
    {\cal X}(\Theta')={\cal X}(\Theta)^{-1},\quad
    {\cal Y}(\Theta')={\cal Y}(\Theta)^{-1}}$
    \\
    $\sigma_x$& 
    $(n',m')=(n,-m)$&
    $\Theta'=-\Theta$&
    $\displaystyle{
    {\cal X}(\Theta')={\cal X}(\Theta),\quad
    {\cal Y}(\Theta')={\cal Y}(\Theta)^{-1}}$
    \\
    $\sigma_y$& 
    $(n',m')=(-n,m)$&
    $\Theta'=-\Theta+\omega_2$&
    $\displaystyle{
    {\cal X}(\Theta')={\cal X}(\Theta)^{-1},\quad
    {\cal Y}(\Theta')={\cal Y}(\Theta)}$
    \\
  \end{tabular}
  \end{ruledtabular}
 \end{table*}
 }

 \newcommand\TABLEtwo{%
 \begin{table*}
  \caption{%
  \label{table:2}
  Conditions for $C_{4v}$ and $C_{6v}$ in addition to those for $C_{2v}$ in \myTable{table:1}.
  In the $C_{4v}$ case, invariance under ${\pi}/{2}$-rotation $c_4$ imposes one more condition between ${\cal X}(\Theta)$ and ${\cal Y}(\Theta)$.
  In the $C_{6v}$ case, invariance under ${\pi}/{3}$-rotation $c_6$ and ${2\pi}/{3}$-rotation $c_6^2$ require two more conditions between ${\cal A}_1(\Theta)$ and ${\cal A}_2(\Theta)$.
  }
  \begin{ruledtabular}
   \begin{tabular}{cllc}
   $C_{4v}$&
   Coordinate trans.&
   Path shifts&
   Condition for ${\cal X}(\Theta)$ and ${\cal Y}(\Theta)$
   \\
    \hline
    $c_4$&
    $(n',m')=(-m,n)$&
    $\Theta'=\Theta+\frac{\omega_2}{2}$&
    ${\cal Y}(\Theta')={\cal X}(\Theta)$
    \\
    \hline
    \hline
    $C_{6v}$&
   Coordinate trans.&
   Path shifts&
   Conditions for ${\cal A}_1(\Theta)$ and ${\cal A}_2(\Theta)$
   \\
   \hline
    $c_6$&
    $(j',i')=(j-i,j)$&
        $\Theta'=\Theta+\frac{\omega_2}{3}$&
    ${\cal A}_1(\Theta'){\cal A}_2(\Theta')={\cal A}_1(\Theta)$
    \\
    $c_6^2$&
    $(j',i')=(-i,j-i)$&
    $\Theta'=\Theta+\frac{2\omega_2}{3}$&
    ${\cal A}_2(\Theta')={\cal A}_1(\Theta)$
    \\
   \end{tabular}
  \end{ruledtabular}
 \end{table*}
 }

\newcommand\TABLEfit{%
\begingroup
\renewcommand{\arraystretch}{0.9}
\begin{table*}
  \caption{%
  \label{table:fit}
  Temperature dependence of the optimal values and ACLs in the row and diagonal directions.
  Annular domains $\myann(C_{\rm max},10^{-7})$ with $C_{\rm max}=10^{-4},~10^{-2},~10^{-4}$ and $5\times10^{-5}$
  were used for $Q=1$, 2, 3, and 4, respectively. 
  Underlined digits in the second part coincide with the exact values and parenthesized numbers are error estimates.
  }
\begin{tabular}{crrllll}
  \hline
  \hline
  $Q$&$~t~~$&$~~|{\cal D}|~~$&$\bar k$&$\bar A$&$\xi_{\rm row}$&$\xi_{\rm diag}$\\
  \hline
  1
  & 0.50&3462~~&0.6414103(7) &1.0331304(205)&3.0054826(63)&3.0054714(63)\\
  & 0.65&2076~~&0.5615632(8) &1.0261963(185)&2.3146357(57)&2.3146114(57)\\
  & 1.00& 966~~&0.4245572(4) &1.0166832(89) &1.5621318(17)&1.5620547(17)\\
  & 1.50& 510~~&0.3036457(1) &1.0100107(5)  &1.1265246(3) &1.1263275(3) \\
  & 2.00& 336~~&0.2299296(5) &1.0066811(124)&0.9164885(13)&0.9161383(13)\\
  & 5.00& 108~~&0.0792507(11)&1.0021472(566)&0.5402944(30)&0.5389299(31)\\
  &14.00&  66~~&0.0195738(5) &0.9999633(50) &0.3571421(2) &0.3538792(2) \\
  \hline
  2
  & 0.20&5994~
  &$\underbar{0.680037}5(4)$ 
  &$\underbar{1.00000}52(55)$
  &$\underbar{3.46041}15(52)$ 
  &$\underbar{3.46040}41(52)$\\
  & 0.30&3042~
  &$\underbar{0.578903}2(6)$ 
  &$\underbar{1.00000}20(83)$
  &$\underbar{2.44298}56(46)$  
  &$\underbar{2.44296}48(46)$\\
  & 0.50&1344~
  &$\underbar{0.438551}8(7)$
  &$\underbar{1.00000}37(57)$   
  &$\underbar{1.62315}82(18)$ 
  &$\underbar{1.62308}92(18)$\\
  & 1.00& 522~
  &$\underbar{0.258819}2(3)$    
  &$\underbar{0.99999}89(43)$   
  &$\underbar{0.995252}1(8)$ 
  &$\underbar{0.994973}2(8)$\\
  & 2.00& 234~
  &$\underbar{0.129039}6(4)$   
  &$\underbar{0.99999}65(64)$
  &$\underbar{0.663502}9(9)$ 
  &$\underbar{0.662678}6(9)$\\
  & 5.00& 114~
  &$\underbar{0.0421156}(0)$   
  &$\underbar{1.000000}3(1)$  
  &$\underbar{0.4373675}(0)$ 
  &$\underbar{0.4351778}(0)$\\
  &10.00&  66~
  &$\underbar{0.016401}8(0)$
  &$\underbar{0.99999}57(78)$ 
  &$\underbar{0.342878}8(5)$
  &$\underbar{0.339369}6(5)$\\
  \hline
  3                                          
  & 0.15&2490~~&0.6267329(4) &0.9615277(102) &2.8568853(19) &2.8568723(19) \\
  & 0.20&2046~~&0.5597793(4) &0.9679882(105) &2.3019859(28) &2.3019611(28) \\
  & 0.30&1164~~&0.4590180(0) &0.9765406(10)  &1.7176157(0)  &1.7175572(0)  \\
  & 0.50& 582~~&0.3320364(4) &0.9856334(119) &1.2166812(13) &1.2165227(13) \\
  & 1.00& 258~~&0.1859817(8) &0.9938120(222) &0.8032343(20) &0.8027337(20) \\
  & 2.00& 144~~&0.0896159(2) &0.9974591(132) &0.5666246(5)  &0.5654062(5)  \\
  & 8.00&  42~~&0.0151040(3) &1.0002672(340) &0.3366384(12) &0.3330153(13) \\
  \hline
  4                                          
  & 0.10&2874~~&0.6265930(2) &0.9176194(7)  &2.8555249(19)  &2.8555119(19) \\ 
  & 0.14&1812~~&0.5552042(22)&0.9321563(549)&2.2699906(152) &2.2699648(152)\\   
  & 0.20&1152~~&0.4739819(12)&0.9468457(269)&1.7909835(60)  &1.7909317(60) \\   
  & 0.30& 684~~&0.3793838(12)&0.9619715(266)&1.3821942(45)  &1.3820842(45) \\   
  & 0.50& 384~~&0.2667541(14)&0.9776521(388)&1.0176226(40)  &1.0173604(40) \\   
  & 1.00& 186~~&0.1449223(5) &0.9915611(105)&0.7020979(12)  &0.7013848(12) \\   
  & 2.00& 108~~&0.0685188(9) &0.9956911(506)&0.5122705(24)  &0.5107258(24) \\   
  & 6.00&  42~~&0.0168894(3) &0.9984632(330)&0.3451569(12)  &0.3416882(12) \\
  \hline
  \hline
  \end{tabular}
  \end{table*}
\endgroup
}

\makeatletter
\renewcommand{\theequation}{%
  \arabic{section}.\arabic{equation}}
\@addtoreset{equation}{section}
\makeatother

\begin{document}
 \title{Universal asymptotic correlation functions for point group $\boldsymbol{C_{6v}}$ and an observation for triangular lattice $\boldsymbol{Q}$-state Potts model}
 \author{Masafumi Fujimoto}
 \affiliation{Department of Physics, Nara Medical University, Kashihara, Nara 634-8521, Japan}
 \author{Hiromi Otsuka}
 \affiliation{Department of Physics, Tokyo Metropolitan University, Hachioji, Tokyo 192-0397, Japan}
 \date{\today}
 \begin{abstract}
 We investigate universal forms for asymptotic correlation functions of off-critical systems that possess $C_{6v}$ symmetry following the argument for $C_{4v}$ symmetry in Phys.~Rev.~E{\bf 102},~032141.
 Unlike the $C_{4v}$ case, a minimal form exists that contains only two free parameters: the normalization constant and modulus.
 Using this form as a building block, we can construct next asymptotic forms to the minimal one. 
 We perform large-scale Monte Carlo simulations of the triangular lattice $Q$-state Potts model above the transition temperature and successfully obtain numerical evidence to support a wide applicability of the minimal form to lattice models, 
 including unsolvable ones.
 From the calculated minimal form, we derive the universal shape of equilibrium crystals 
 in the honeycomb lattice Potts model described by an algebraic curve of genus 1.
 Although the curve differs from those obtained in the $C_{4v}$ case, 
 the latters also have genus 1. 
 We indicate that the birational equivalence concept can play an important role in comparing asymptotic forms for different point group symmetries, for example, $C_{6v}$ and $C_{4v}$.
 \end{abstract}

 \maketitle

 \newpage
 \section{INTRODUCTION}
 \label{sec:1}
 The thermal evolution of the equilibrium crystal shape (ECS) is a long-standing problem. 
 In 1901, Wulff \cite{Wulff1901} proposed a method to determine the ECS from the anisotropic interfacial tension, which is called Wulff's construction; also see \cite{Laue1944,Herring1951}. 
 ECSs are polygons or polyhedrons at the zero-temperature limit.
 They become circles or spheres near the critical temperature 
 when the interactions are isotropic. 
A roughening transition exists between them \cite{Burton1951}. 

 The roughening transition should be investigated as a cooperative phenomenon from the microscopic Hamiltonian within the framework of statistical mechanics. 
 Practically, however, it is quite difficult for these calculations to determine reliable results.
 In the 1970s, the development of exact analyses of solvable lattice models enabled detailed studies of the roughening transition. 
 In the early stage of research, the disappearance of a sharply defined interface at the roughening transition temperature $T_{\rm R}$ received much attention. 
 In \myRefs{Abraham1974,Abraham1976}, the interface profile of the square lattice Ising model was investigated to show that its width diverges at the thermodynamic limit. 
 Then, it was recognized that this phase transition may influence crystal morphology \cite{Beijeren1977,Jayaprakash1983,Rottman1981,Avron1982,Zia1982}. 
 Pioneering research on this issue was conducted in \myRefs{Beijeren1977,Jayaprakash1983}, where the body-centered solid-on-solid (BCSOS) model was investigated with the help of the exact solutions of the six-vertex model \cite{Lieb1972,Baxter2007}. 
 The facet shape in the BCSOS model was analyzed to identify a universal jump at $T_{\rm R}$ in the principal curvature of the two-dimensional (2D) surface of the three-dimensional ECS; also see \myRefs{Landau1980,Andreev1981}.
 On the other hand, for the square lattice Ising model, the 2D ECS was calculated in \myRefs{Rottman1981,Avron1982,Zia1982} and expressed as a simple algebraic curve in the $\alpha\beta$ plane: 
 \begin{equation}
  \alpha^2\beta^2+1+A_3(\alpha^2+\beta^2)+A_4\alpha\beta=0,
   \label{eq:1.1}
 \end{equation}
 with $\alpha=\exp[-\lambda(X+Y)/\mykB T]$ and $\beta=\exp[-\lambda(X-Y)/\mykB T]$, where $(X,Y)$ is the position vector of a point on the ECS and $\lambda$ is a scale factor; for the definitions of $A_3$ and $A_4$, see \myRef{Zia1982}. 
 
 In \myRef{Akutsu1990}, the authors indicated that the ECS (\ref{eq:1.1}) is identical to the facet shape of the BCSOS model. 
 The accumulation of research on ECSs revealed a paradoxical scenario:
 Clearly, the interfaces of lattice models have model-dependent microscopic profiles (see, e.g., \myRef{Selke1982a}).  
 \myEquation{eq:1.1} commonly represents the ECSs of a wide range of solvable models \cite{Fujimoto1992,Fujimoto1993,Fujimoto1997}. 
 Furthermore, for some unsolvable models, researchers showed that the ECS (or facet shape) is quite close to \myEq{eq:1.1} \cite{Akutsu1987a,Akutsu1987,Holzer1989}. 

 For the square lattice Ising model, the anisotropic correlation length (ACL) is related by duality to the anisotropic interfacial tension \cite{Zia1978}. 
 The ECS (\ref{eq:1.1}) was derived from the ACL via the duality relation and  Wulff's construction; also see \myRefs{Akutsu1990,Holzer1990,Holzer1990a}. 
 Thus, \myEq{eq:1.1} appears in the representation of the asymptotic correlation function of the square lattice Ising model.
 The same scenario was found in other solvable models on the square lattice without the duality relation \cite{Fujimoto1992,Fujimoto1993}.
 These facts suggested a close relation between the algebraic curve (\ref{eq:1.1}) and 
 symmetry properties 
 of the square lattice; see, for example, \myRef{Hamermesh1989}.

 Quite recently, Fujimoto and Otsuka \cite{Fujimoto2020} investigated the asymptotic correlation functions of the square lattice $Q$-state Potts model.  
 The model is solvable at the phase transition point \cite{Baxter2007,Temperley1971,Baxter1973,Wu1982}.
 For $Q>4$, the phase transition is first order. 
 Researchers showed that \myEq{eq:1.1} appears in the asymptotic behavior of the correlation function  at the first-order transition point \cite{Fujimoto1997}; also see \myRef{Kluemper1989,Buffenoir1993}. 
 When $Q=2$, the Potts model reduces to the Ising model.
 We reexamined its asymptotic correlation function both above and below the transition temperature.
 Using the combination of the transfer matrix and shift operator \cite{Fujimoto1990,Fujimoto1990a,Fujimoto1992}, we reproduced the same results as those using the Pfaffian method \cite{Cheng1967,McCoy2013}; also see \myRef{Yamada1983,Yamada1984,Yamada1984a,Yamada1986}. 
 Furthermore, we analyzed the Ising model on a square lattice rotated through an arbitrary angle with respect to the coordinate axes. 
 Johnson, Krinsky, and McCoy \cite{Johnson1973} showed that the summation over the eigenvalues of the transfer matrix becomes contour integrals in the thermodynamic limit.  
 Although lattice rotation causes the integration paths to move, the contour integrals must be independent of the path movement. 
 We found that (i) the  analyticity of the eigenvalues of the transfer matrix and the shift operator is necessary to ensure path independence with the help of Cauchy's theorem; 
 (ii) satisfying a functional equation corresponds to $\pi$-rotational invariance; and because $2\pi$ rotation returns the contour integrals to the original integrals, (iii) the eigenvalues possess doubly periodic structures.
 
 The three properties (i)--(iii) directly connect with $C_{2v}$ symmetry, not with the exact solvability of the Ising model; hence, they apply to a general $Q$.
 Using these properties, we obtained the general asymptotic form \exEq{(2.1)} with (2.2) in \myRef{Fujimoto2020}. 
 The system possesses $C_{4v}$ symmetry when the interactions are isotropic. 
 Because $C_{2v}$ is a normal subgroup of $C_{4v}$, we found \exEq{(2.3)} [or equivalently \exEq{(3.2)}] in \myRef{Fujimoto2020}, which contains three free parameters. 
 Regarding them as fitting parameters, we numerically analyzed correlation function data provided by Monte Carlo (MC) simulations. 
 Then we obtained strong evidence that the asymptotic form 
 \exEq{(3.2)} in \myRef{Fujimoto2020} applies to the correlation function and the ACL in the disordered phase. 
 From the calculated ACLs, we derived the ECSs via duality \cite{Laanait1987,Fujimoto1997} and Wulff's construction. 
 Then, we found that \myEq{eq:1.1} should be replaced by
 \begin{equation}
  \alpha^2\beta^2
   +
   1
   +
   {\bar A}_2(\alpha\beta+1)(\alpha+\beta)
   +
   \alpha^2+\beta^2
   +
   {\bar A}_4\alpha\beta
   =
   0
   \label{eq:1.2}
 \end{equation}
 (also see \mySec{sec:4.B} and \myRef{Fujimoto1996}).
 We successfully detected a small deformation in the ECS: 
 In \exEq{(3.2)} in \myRef{Fujimoto2020}, one of the free parameters was denoted by $b$.
 For $Q=2$, $b=1$, that is, an exact value, where \myEq{eq:1.2} is reduced to \myEq{eq:1.1}
 \cite{Fujimoto1996}. 
 The MC data showed that $b>1$ for $Q=1$ (the bond percolation), and $b<1$ for $Q=3$, 4. 
 According to \myEq{eq:1.2}, the ECS is rounded in the facet directions and flattened in the corner directions as $Q$ increases. 

 In this paper, following the analysis for the $C_{4v}$ case \cite{Fujimoto2020}, we consider asymptotic correlation functions with $C_{6v}$ symmetry.
 Numerous attempts have been made to calculate the asymptotic correlation functions on the triangular lattice Ising model where the Pfaffian method was used \cite{Vaidya1976}; also see \myRefs{Stephenson1964,Chan2011} and \exChap{VIII} of \myRef{McCoy2013}. 
 Furthermore, the asymptotic form of the triangular/honeycomb lattice Potts model 
 at the first-order transition point \cite{Fujimoto1999} 
 and that in the Kagom\'{e} lattice eight-vertex model \cite{Fujimoto2002} were 
 calculated using the transfer matrix argument. 
 However, little is known about the $C_{6v}$ symmetry in asymptotic correlation functions.
 
 Our strategy for investigating the $C_{6v}$ symmetric case is as follows:
 We expect that the transfer matrices satisfy the three properties (i)--(iii) mentioned above; 
 note that $C_{6v}$ also contains the normal subgroup $C_{2v}$. 
 This leads us to 
 the asymptotic correlation function given by \exEq{(2.1)} with (2.2) in \myRef{Fujimoto2020}. 
 The factor group $C_{4v}/C_{2v}$ is the cyclic group of order 2. 
 On the other hand, the factor group $C_{6v}/C_{2v}$ is the cyclic group of order 3. 
 As discussed below, we find a minimal form that includes only two parameters. 
 This is because one more condition than that for $C_{4v}$ fixes the $b$ parameter and then yields a model-independent minimal form for $C_{6v}$. 
 Although the number of free parameters, that is, two, is equal to that included in the so-called 
 Ornstein--Zernike (OZ) form, our minimal form possesses discrete $C_{6v}$ symmetry.

 To proffer numerical evidence to support the applicability of the minimal form with $C_{6v}$, we perform large-scale MC simulations of the triangular lattice $Q$-state Potts model \cite{Baxter2007,Wu1982}. 
 We analyze the MC data of the asymptotic correlation functions for the $Q=1$, 2, 3, and 4 cases above critical temperature. 
 We find that the minimal form well fits numerical data in these cases and yields precise estimates of the ACLs.
 Additionally, we fit the data  using the OZ form and reveal the superiority of the minimal form via a comparison of their fittings. 

 We present some implications of our findings for asymptotic correlation functions on the triangular lattice. 
 Unlike the case of $C_{4v}$, ACLs include only one parameter: the modulus. 
 This fact means that, in discussing long-distance behavior, Potts models with various $Q$ have the same character, with a mere re-scaling of the temperature. 
 Using the ACLs obtained from the minimal form, we derive the ECS on the honeycomb lattice via the duality transformation and Wulff's construction, which suggests that the same asymptotic forms appear in the honeycomb lattice Potts model. 
 Furthermore, similar to the case of $C_{4v}$ symmetry, the ECS is given by an algebraic curve of genus 1. 
 Based on the previous study \cite{Fujimoto2020} and present study, we can explain the mathematical background to relate the correlation functions of different models, even on different 2D lattices. 

 The present paper is organized as follows:
 In \mySec{sec:2}, we provide the minimal form adaptable to asymptotic correlation functions with $C_{6v}$ symmetry. 
 In \mySec{sec:3}, we perform MC simulations of the triangular lattice $Q$-state Potts model in disordered phases.
 The minimal form fits the numerical data of correlation functions well and yields precise estimates of the ACLs.
 Additionally, we fit data using the OZ form and reveal a superiority of the minimal form via a comparison of their fittings. 
 In \mySec{sec:4}, we discuss and summarize the study.
 We derive an ECS on the honeycomb lattice from the ACLs in triangular lattice models. 
 Then we provide a birational transformation  \cite{Walker1950} to connect the algebraic curve for $C_{6v}$ to that for $C_{4v}$. 
 In \myAppendix{appendix:A} and \myAppendix{appendix:B}, we explain the exact calculation of the asymptotic correlation function of the triangular lattice Ising model and derive the birational transformation given in \mySec{sec:4}, respectively. 
 \newpage
 \section{ASYMPTOTIC CORRELATION FUNCTIONS FOR $\boldsymbol{C_{6v}}$}
 \label{sec:2}
 To make our discussion specific, we assume a triangular lattice on which the $Q$-state Potts model with isotropic interactions is defined. 
 The exact calculation of the $Q=2$ Potts model explicitly shows that the three properties (i)--(iii) are satisfied \cite{Fujimoto2002,Fujimoto2020}.
 We derive general forms of the correlation functions with $C_{6v}$ symmetry using them as necessary basic conditions.

 As depicted in \myFig{fig:1}, a triangular lattice consists of all points with position vectors ${\bf r}=j{\bf a}_1+i{\bf a}_2$, where the primitive vectors are denoted by ${\bf a}_1$ and ${\bf a}_2$. The lattice spacing $|{\bf a}_1|=|{\bf a}_2|=a$ and the angle between them is $2\pi/3$. 
 For the triangular lattice Ising model, the asymptotic correlation function was analyzed using the Pfaffian method \cite{Vaidya1976}; also see \myRefs{Holzer1990a,Zia1986}.
 We can derive essentially the same results by introducing the shift operator into the usual transfer matrix method. 
 We restrict ourselves to the case of isotropic interactions, where the system possesses $C_{6v}$ symmetry.
 To find the role of $C_{6v}$ symmetry, we investigate the model 
 on triangular lattices rotated clockwise through various angles to the coordinate axes: we consider the cases of the rotation angle $n\pi/6$ with $n=0,1,\dots,11$.
 We summarize the main results in the main text and provide the details in \myAppendix{appendix:A}.

 In the thermodynamic limit, the summations over eigenvalues of the transfer matrix (and those of the shift operator) become integrals because of their continuous distribution.
 The asymptotic correlation functions are represented using contour integrals on Riemann surfaces, as shown in \myEqs{eq:A8} and (\ref{eq:A9}); also see \cite{Baxter2007,Johnson1973}. 
 Considering transfer matrices in the rotated systems shows that the three properties (i)--(iii) are fulfilled:
 In \myAppendix{appendix:A2}, we indicate that the lattice rotations shift the integration paths. 
 After the eigenvalues are summarized, thermodynamic averages must be independent of the rotation angle.
 This equivalence is derived with the help of the analyticity of the eigenvalues of the transfer matrices and the shift operators.
 Hence, (i) the analyticity of the eigenvalues is indispensable; 
 (ii) the eigenvalues should satisfy a functional equation corresponding to $\pi$-rotational symmetry; see \exEq{(A27)} or \exEq{(A28)} in \myRef{Fujimoto2020}; and (iii) the eigenvalues should possess two periods. 
 Intuitively, we can explain this property as follows:
 The 2D lattice models are related to the 2D Euclidean field theories in their critical limit and for distances much larger than $a$. 
 The correlation function has the periodicity of rotational symmetry in the limit.
 For off-critical lattice models, the crystal momentum is defined as modulo ${2\pi}/{a}$. 
 Therefore, lattice models possess two types of periodicity: 
 one is the two, four, or sixfold rotational symmetry, and the other is that the eigenvalues of transfer matrices are periodic functions of the crystal momentum.

 \FIGone
 \FIGtwo

 Because $C_{2v}$ is the normal subgroup of $C_{6v}$, we consider $C_{2v}$ symmetry as the first step;
 that is, we start with (i)--(iii) to shape the asymptotic correlation functions with $C_{6v}$ symmetry.
 For this purpose, it is convenient to divide the triangular lattice into two sublattices shown by open and closed circles in \myFig{fig:1}.
 Let ${\bf o}$ be the position vector of the origin, and ${\bf r}^+$ that of another site in the same sublattice.
 Then, in terms of the primitive vectors of sublattice ${\bf b}_1$ and ${\bf b}_2$, ${\bf r}^+=n{\bf b}_1+m{\bf b}_2=(n+m){\bf a}_1+2m{\bf a}_2$.
 
 Because of property (iii), by choosing a suitable parameterization, we can represent the asymptotic correlation function between $\bf o$ and ${\bf r}^+$ as a contour integral on a Riemann surface of genus 1 (see \myFig{fig:2}):
 \begin{equation}
  {\cal F}_{{\bf o},{\bf r^+}}
  \sim{\rm const}
   \int^{\omega_1}_{-\omega_1}~d\Theta~
   {\cal X}(\Theta)^n
   {\cal Y}(\Theta)^m,
  \label{eq:2.1}
 \end{equation}
 where ${\cal Y}(\Theta)$ corresponds to eigenvalues of the row-to-row transfer matrix along the vertical direction and ${\cal X}(\Theta)$ to those of the shift operator along the horizontal direction. 
 As explained above,
 ${\cal X}(\Theta)$ and ${\cal Y}(\Theta)$ are doubly periodic functions: 
 ${\cal X}(\Theta+2\omega_1)={\cal X}(\Theta+2\omega_2)={\cal X}(\Theta)$
 and 
 ${\cal Y}(\Theta+2\omega_1)={\cal Y}(\Theta+2\omega_2)={\cal Y}(\Theta)$.
 Then, according to property (ii), invariance under $\pi$-rotation (say $c_2$) enforces the following functional equations on the eigenvalues:
 \begin{equation}
 {\cal X}(\Theta+\omega_2)={\cal X}(\Theta)^{-1},
 \quad
 {\cal Y}(\Theta+\omega_2)={\cal Y}(\Theta)^{-1}.
 \label{eq:2.2}
 \end{equation}
  From (i), we assume suitable analytic properties for ${\cal X}(\Theta)$ and ${\cal Y}(\Theta)$, 
  and then, using their series expansions, obtain  
 \begin{equation}
 {\cal X}(\Theta)
 =
 \prod^{\nu}_{l=1}k^{\frac12}\mySn{\Theta+\alpha_l},
 \quad
 {\cal Y}(\Theta)
 =
 \prod^{\nu'}_{l=1}k^{\frac12}\mySn{\Theta+v+\beta_l},
 \label{eq:2.3}
 \end{equation}
 where $k\in(0,1)$ is the modulus corresponding to the modular parameter 
 $\tau=\omega_2/\omega_1$; see \exApp{A.3} of \myRef{Fujimoto2020}.
 For definitions of Jacobi's elliptic functions, see \exChap{15} of \myRef{Baxter2007}.

 \TABLEone

 In addition to $c_2$, it is necessary to consider the invariance of \myEq{eq:2.1} under the vertical reflection ($\sigma_x$) or the horizontal reflection ($\sigma_y$) to achieve $C_{2v}$ symmetry.
 For example, we obtain the following functional equations from invariance under $\sigma_x$:
 \begin{equation}
 {\cal X}(-\Theta)={\cal X}(\Theta),
 \quad
 {\cal Y}(-\Theta)={\cal Y}(\Theta)^{-1}. 
 \label{eq:2.4}
 \end{equation}
 Because $\sigma_y=\sigma_x\cdot c_2$, the conditions~(\ref{eq:2.2}) and (\ref{eq:2.4}) yield the invariance of \myEq{eq:2.1} under $\sigma_y$.
 In \myTable{table:1}, we summarize the functional equations to achieve $C_{2v}$ symmetry.
 Because $\mySn{-\Theta}=-\mySn{\Theta}$, we find that $v=-\omega_2/2$, and $\nu$ and $\nu'$ are even integers.
 Additionally, $\tau$ must be purely imaginary because the correlation function is real-valued (see below). 

 It should be noted that $c_2$ shifts integration paths by $\omega_2$ without deforming them.
 In this sense, twofold rotational symmetry divides a periodic rectangle into two sub-regions.
 Meanwhile, the reflections $\sigma_x$ and $\sigma_y$ cause reflections of paths about the blue and red dotted lines in \myFig{fig:2}(a), respectively.
 Consequently, the equivalent integration paths appear repeatedly in the periodic rectangle because of $C_{2v}$ symmetry.
 
 \TABLEtwo

 We showed \cite{Fujimoto2020} that to derive the asymptotic forms for $C_{4v}$ from those for $C_{2v}$, fourfold rotational symmetry requires an additional functional equation: ${\cal Y}(\Theta)={\cal X}(\Theta-{\omega_2}/2)$ (see the upper part of \myTable{table:2}).
 Similarly, we construct an asymptotic form for $C_{6v}$ from those for $C_{2v}$. 
 To achieve this, it is convenient to introduce ${\cal A}_1(\Theta)$ and ${\cal A}_2(\Theta)$, which are associated with primitive translations of ${\bf a}_1$ and ${\bf a}_2$, respectively.
 These are related to ${\cal X}(\Theta)$ and ${\cal Y}(\Theta)$ as
 \begin{equation}
 {\cal X}(\Theta)={\cal A}_1(\Theta),
 \quad
 {\cal Y}(\Theta)={\cal A}_2(\Theta)^2{\cal A}_1(\Theta)
 \label{eq:2.5}
 \end{equation}
 (see \myFig{fig:1}).
 Then, sixfold rotational symmetry yields two additional functional equations:
 \begin{equation}
 {\cal A}_2(\Theta)
 =
 {\cal A}_1\left(\Theta-\frac{2\omega_2}3\right),
 \quad
 {\cal A}_1(\Theta){\cal A}_2(\Theta)
 =
 {\cal A}_1\left(\Theta-\frac{\omega_2}3\right)
 \label{eq:2.6}
 \end{equation}
  (see the lower part of \myTable{table:2}).
 Using ${\cal A}_1(\Theta)$ and ${\cal A}_2(\Theta)$, we can express the correlation function between ${\bf o}$ and ${\bf r}=j{\bf a}_1+i{\bf a}_2$ as
 \begin{equation}
 {\cal F}_{{\bf o},{\bf r}}
  \sim{\rm const}
   \int^{\omega_1}_{-\omega_1}~d\Theta~
   {\cal A}_1(\Theta)^j
   {\cal A}_2(\Theta)^i.
 \label{eq:2.7}
 \end{equation}

 \subsection{Minimal case for $\boldsymbol{C_{6v}}$}
 \label{sec:2.A}
 
 The unit cell of a sublattice is vertically long (see \myFig{fig:1}), and the relations~(\ref{eq:2.5}) and (\ref{eq:2.6}) require a condition for the integers: $\nu'=2\nu$.
 Therefore, we can find the simplest expression by setting $\nu=2$ and $\nu'=4$ in \myEq{eq:2.3}, whose parameters are fixed as
 $\alpha_1=-\alpha_2=\omega_2/6$, $\beta_1=\beta_2=0$, and $\beta_3=-\beta_4=\omega_2/3$.
 As a result, a minimal form of the asymptotic correlation function with $C_{6v}$ symmetry is given by
 \begin{align}
 &{\cal F}^{\text{(min)}}_{{\bf o},{\bf r}}
 =
 {\rm const}
 \int^{\omega_1}_{-\omega_1}~d\Theta
 \cr
&\mySnSn{\Theta}{+\frac{\omega_2}{6}}{-\frac{\omega_2}{6}}^j
 \mySnSn{\Theta}{-\frac{\omega_2}{2}}{-\frac{5\omega_2}{6}}^i.
 \label{eq:2.8}
 \end{align}
 The integrand is built from two elliptic functions, each composed of two sn functions, and possesses essentially the same structure as the simplest case for $C_{4v}$ \cite{Fujimoto2020}.
 As different points, we replace the value of $\omega_2/2$ in \exEq{(2.3)} in \myRef{Fujimoto2020} with $2\omega_2/3$, and fix the undetermined constant $B$ to $\omega_2/6$ because $C_{6v}$ requires not one but two additional conditions, as given in \myTable{table:2}.
 Consequently, \myEq{eq:2.8} contains only two parameters: a normalization constant and modulus $k$.

 Once we determine the expressions of these parameters as \myEqs{eq:A8a} and (\ref{eq:A8b}), \myEq{eq:2.8} provides the leading asymptotic correlation function of the triangular lattice Ising model above the critical temperature $T>\myTc$. 
 Because the pair of integers $(\nu,\nu')=(2,4)$ cannot change as a result of continuous variations of $Q$, the minimal case applies unless a phase transition occurs. 
 Indeed, we found that the simplest case with $\nu=\nu'=2$ is commonly observed in the square lattice $Q$-state Potts model in the disordered phase \cite{Fujimoto2020}. 
 Therefore, we expect that the minimal form (\ref{eq:2.8}), including the normalization factor and modulus as free parameters, describes the leading asymptotic behavior of the triangular lattice $Q$-state Potts model above the transition temperature $\myTc(Q)$ [see \myEq{eq:3.2}]. 

 \subsection{Next to minimal case}
 \label{sec:2.B}
 
 We obtain the next to minimal case by setting $(\nu,\nu')=(4,8)$ in \myEq{eq:2.3}.
 In addition to $\alpha_1,~\alpha_2$, $\beta_1,\dots\beta_4$, six parameters exist; say $\bar{\alpha}_1,~\bar{\alpha}_2$, $\bar{\beta}_1$, $\dots\bar{\beta}_4$.
 We introduce ${\bar\Theta}$ for $\bar{\alpha}_i$s and $\bar{\beta}_j$s, which we determine by repeating the same argument as that in the previous subsection. 
 Then, we obtain a form for the next to minimal case as follows:
 \begin{align}
 &{\cal F}^{\rm (next)}_{{\bf o},{\bf r}}
 =
 \int^{\omega_1}_{-\omega_1}~d\Theta
 \int^{\omega_1}_{-\omega_1}~d{\bar\Theta}\ \rho(\Theta-{\bar \Theta})\cr
 \times
&\mySnSn{\Theta}{+\frac{\omega_2}{6}}{-\frac{\omega_2}{6}}^j
 \mySnSn{\Theta}{-\frac{\omega_2}{2}}{-\frac{5\omega_2}{6}}^i
 \cr
 \times
&\mySnSn{\bar\Theta}{+\frac{\omega_2}{6}}{-\frac{\omega_2}{6}}^j
 \mySnSn{\bar\Theta}{-\frac{\omega_2}{2}}{-\frac{5\omega_2}{6}}^i,
 \label{eq:2.9}
 \end{align}
 where $\rho(\Theta)$ is an even function that we determine from the distribution of the eigenvalues and the matrix elements \cite{Johnson1973}.
 Equation~(\ref{eq:2.9}) indicates that the integral on the right-hand side of \myEq{eq:2.8} plays the role of a building block of correlation functions.
 Because \myEq{eq:2.9} is a higher-order term of the pairs of sn functions, it is naturally regarded as a correction to \myEq{eq:2.8} for the triangular lattice Potts model above $\myTc(Q)$.

 The scenario becomes somewhat complicated below $\myTc(Q)$: 
 In the triangular lattice Ising model, the leading asymptotic behavior of the correlation function is given by \myEq{eq:2.9}; see \myAppendix{appendix:A}. 
 The same is expected in the triangular lattice $Q$-state Potts model.
 Despite this, we discuss the possibility of observing the minimal case~(\ref{eq:2.8}) below $\myTc(Q)$ in \mySec{sec:4}.

 \newpage
 \section{NUMERICAL ANALYSES OF TRIANGULAR LATTICE $\boldsymbol{Q}$-STATE POTTS MODEL}
 \label{sec:3}
 The Hamiltonian of the triangular lattice $Q$-state Potts model is given by
 \begin{equation}
 H
 =
 -J\sum_{\langle{\bf r,r'}\rangle}
 \left[2\delta(q_{\bf r},q_{\bf r'})-1\right]
 \quad
 (J>0),
 \label{eq:3.1}
 \end{equation}
 where the $Q$-valued variable $q_{\bf r}=0,1,\dots,Q-1$ is associated with a site in a triangular lattice ${\bf r}\in\Lambda_{\rm tri}$ and the sum runs over all nearest-neighbor pairs of sites.
 For each $q_{\bf r}$, we introduce a spin variable $\sigma_{\bf r}=\exp(2\pi\myIm q_{\bf r}/Q)$. 

 The phase transition is continuous for $Q\leq 4$ and first order for $Q>4$ \cite{Wu1982,Baxter2007}. The transition temperature $\myTc(Q)$ is given by
 \begin{equation}
 \sqrt{Q} x_{\rm C}^3+3x_{\rm C}^2=1,
 \quad 
 \sqrt{Q} x_{\rm C}=e^{2J/k_{\rm B}\myTc(Q)}-1.
 \label{eq:3.2}
 \end{equation}
 In this section, we restrict ourselves to the disordered phase: $T>\myTc(Q)$.  
 The spin correlation function $c({\bf r})$ is defined as
 \begin{equation}
 c({\bf r-r'})
 =
 \langle \sigma_{\bf r}\sigma_{\bf r'}^{\ast}\rangle
 =
 \myav{\exp\left[\frac{2\pi\myIm (q_{\bf r}-q_{\bf r'})}{Q}
 \right]}.
 \label{eq:3.3}
 \end{equation}

 Some MC algorithms are known to simulate Potts models efficiently \cite{Swendsen1987, Wolff1988,Wolff1989}. 
 In the previous study of the square lattice, $Q$-state Potts model \cite{Fujimoto2020},
 we used a cluster MC algorithm for infinite-size systems proposed by Evertz and von~der~Linden \cite{Evertz2001}.
 It allows us to simulate off-critical systems in the thermodynamic limit directly.
 We can also benefit from these strong points in the numerical analysis of \myEq{eq:3.1}, which provides solid ground to check the applicability of the $C_{6v}$ form for asymptotic correlation functions.

 First, we summarize the methodological aspect of MC simulations by borrowing some notation and definitions provided in \myRef{Fujimoto2020}. 
 The MC algorithm is based on the Fortuin--Kasteleyn representation of the partition function of \myEq{eq:3.1}, say $Z(Q)$ \cite{Fortuin1972}.
 Suppose $n_{\bf r^*}~(=0,1)$ is an occupation number of a site ${\bf r^*}$ in the medial lattice of $\Lambda_{\rm tri}$. Then $Z(Q)$ represents a bond percolation on $\Lambda_{\rm tri}$ with the percolation probability $p(T)=1-e^{-\frac{J}{\mykB T}}$.
 Each cluster generated in the percolation process randomly possesses a $Q$-valued color property.
 Therefore, the $Q\to1$ limit of the Potts model provides the standard bond percolation defined on $\Lambda_{\rm tri}$.
 
 The infinite-system MC method \cite{Evertz2001} is based on Wolff's single-cluster algorithm \cite{Wolff1989}. 
 It fixes the seed site to the origin of a lattice ${\bf o}$ throughout a simulation.
 For disordered systems with correlation length $\xi$, we start with random spin configurations on finite lattice systems with linear dimension $l_{\rm B}$.
 The initial MC steps are equilibrating spin configurations within a circular domain that gradually broadens toward its outer region.
 We denote the ratios of an equilibrated circular domain as $l_{\rm T}$, and the number of MC steps required increases exponentially as $\exp(a l_{\rm T}/\xi)$.
 Typically, we prepare equilibrated spin configurations with $l_{\rm T}\simeq20~\!\xi$, and then calculate the MC averages of the correlation functions within the circular domains.
 We use finite systems that satisfy the condition $l_{\rm B}\gg l_{\rm T}$ (in a typical case $l_{\rm B}\simeq 4~\!l_{\rm T}$). 
 Then, the probability of the generated clusters touching the lattice boundary is negligible during viable MC simulation steps.
 
 Because of the random cluster representation of the Potts model, the so-called improved estimator for the correlation functions is available: 
 Suppose ${\cal C}\subset\Lambda_{\rm tri}$ is a set of sites that form a cluster. Then we evaluate the spin correlation functions as
 \begin{equation}
 c({\bf r-r'})
 =
 \myav{\frac{Q\delta(q_{\bf r},q_{\bf r'})-1}{Q-1}}
 =
 \myav{\frac1{|{\cal C}|}\delta({\bf r,r'}~\!|~\!{\cal C})}_{\!\!\rm MC}, 
 \label{eq:3.4}
 \end{equation}
where $|{\cal C}|$ is the number of sites in ${\cal C}$, and $\delta({\bf r,r'}~\!|~\!{\cal C})=1$ if ${\bf r, r'}\in{\cal C}$, and $\delta({\bf r,r'}~\!|~\!{\cal C})=0$ otherwise.
 For $Q=1$, 2, 3, and 4 and at several reduced temperatures $t=[T-\myTc(Q)]/\myTc(Q)$, we prepare the correlation function data and associated statistical errors as functions of ${\bf r}=j{\bf a}_1+i{\bf a}_2$, say $\{c(i,j), d(i,j)\}$.
 In this step, we typically generate about $6\times10^{15}$ clusters for each average calculation to satisfy a high statistical accuracy requirement (see below).

 Now we check the applicability of \myEq{eq:2.8} for the triangular lattice $Q$-state Potts model. 
 For convenience, we replace $\omega_1$ and $\omega_2$ with $I$ and $\myIm I'$, respectively.
 Then, the form for $C_{6v}$ is given by
 \begin{align}
  &{\cal F}_{\rm tri}(i,j;A,k)
  =
   \frac{A(1-k^2)^{\frac14}}{\pi}
   \int^I_{-I}d\varphi
   \cr
  &\mySnSn{\varphi}{+\frac{\myIm I'}{6}}{-\frac{ \myIm I'}{6}}^j
   \mySnSn{\varphi}{-\frac{\myIm I'}{2}}{-\frac{5\myIm I'}{6}}^i.
  \label{eq:3.5}
 \end{align}
 We represent the normalization factor using a parameter $A$ \cite{Fujimoto2020}, which refers to the exact value $A=1$ for $Q=2$; see \myEq{eq:A8b}.
 We use the reduced chi-square statistic to fit the calculation of the $C_{6v}$ form (\ref{eq:3.5}) for MC data, and then extract optimal values by minimizing $\chi^2_{\rm tri}(A,k)$ concerning $A$ and $k$: 
 \begin{equation}
   \chi^2_{\rm tri}(A,k)
   =
   \sum_{(i,j)\in\myann}
   \left[
   \frac{{\cal F}_{\rm tri}(i,j;A,k)-c(i,j)}{d(i,j)}
   \right]^2.
  \label{eq:3.6}
 \end{equation}
 
 In this process, we should pay attention to a region $\myann$ in which the form will fit the MC data. 
 There are two types of sources of errors in the fitting calculations: statistical and systematic errors.
 \myEquation{eq:3.5} does not take the contributions of eigenvalues with $(\nu,\nu')\ne(2,4)$ into account (see \myAppendix{appendix:A}), which causes a systematic error for short-distance fittings.
 By contrast, the longer the distances, the larger the statistical error of MC data, which causes uncertainty in the estimated optimal values.
 As discussed in \myRef{Fujimoto2020}, to control the two types of errors, we use an annular region defined as
 $\myann(c_{\rm max},c_{\rm min})= \{(i,j)\!~|~\!c_{\rm min}<c(i,j)<c_{\rm max}\}$
 and check the $\myann$ dependence of the fitting conditions.
 
 We determine a lower cut-off $c_{\rm min}$ so that the fitting is not affected by a statistical error.
 As mentioned above, we perform large-scale MC calculations, which allow us to use a small value independently of $Q$, for example, $c_{\rm min}=10^{-7}$.
 We should determine the upper cut-off $c_{\rm max}$ according to the magnitude of systematic errors.
 For $Q=2$, the correction from the second-band eigenvalues is absent because of $\mathbb{Z}_2$ symmetry; see \myAppendix{appendix:A} and \myRef{Fujimoto2020}. 
 However, for $Q\ne2$, it does exist in the MC data.
 Therefore, we use $c_{\rm max}$ depending on $Q$; see below.

 Once we obtain the optimal values, say ${\bar A}$ and ${\bar k}$, the correlation function is asymptotically given by 
 $c({\bf r})\sim{\bar{\cal F}}_{\rm tri}(i,j)={\cal F}_{\rm tri}(i,j;{\bar A},{\bar k})$.
 From this expression, we can find the ACLs for the $Q$-state Potts model; see \mySec{sec:4.A}.

 \TABLEfit

 \subsection{Potts model with $\boldsymbol{Q=2}$}
 \label{sec:3.A}
 
 First, we analyze the triangular lattice Ising model because the exact results are available for checking the accuracy of our numerical procedure.

 The second part of \myTable{table:fit} summarizes the fitting results of $Q=2$.
 The geometrical properties of $\myann$ and the ACLs in the row and diagonal directions are given for several reduced temperatures $t$, where $|\myann|$ denotes the number of sites in $\myann$.
 Concerning $c_{\rm max}$, we observe that the fitting condition is almost independent of it, and
 thus use a larger value, that is, $c_{\rm max}=10^{-2}$, to improve statistical accuracy. 
 
 Then we find that the optimized parameters agree well with the exact values, that is, at all temperatures, they yield $\bar A=A_{\rm exact}=1$ and $\bar k=k_{\rm exact}$ with at least six-digit accuracy.
 In the table, note that the underlined digits coincide with the exact values and the parenthesized digits are error estimates.
 As $t$ decreases, the directional dependence of the correlation length becomes weaker; hence, highly accurate numerical data are necessary for its detection.
 The second part of \myTable{table:fit} shows that our numerical approach using the form (\ref{eq:3.5}) is sufficiently efficient to analyze the ACLs with $C_{6v}$ symmetry (see $\xi_{\rm row}$ and $\xi_{\rm diag}$).  
 
 In our previous paper \cite{Fujimoto2020}, we proposed the $C_{4v}$ form and established its goodness of fit for correlation functions of the square lattice Potts model. 
 Naturally, we expect that the advantages explained contribute to the present high accuracy. 
 Additionally, as given in \mySec{sec:2}, $C_{6v}$ symmetry reduces the number of free parameters in the form to two. 
 Therefore, we recognize that the triangular lattice offers a more suitable framework for studying the directional dependence of correlation functions.

 \subsection{Potts model with $\boldsymbol{Q\ne2}$}
 \label{sec:3.B}

 Next, we investigate the applicability of \myEq{eq:3.5} to the triangular lattice $Q=1$, 3, and 4 Potts model.
 In the $C_{4v}$ case, the deformation parameter, $b$ (=1 for the Ising case), exists and represents the $Q$ dependence of the ACLs; see \exSec{3} of \myRef{Fujimoto2020}.
 By contrast, \myEq{eq:3.5} only includes amplitude $A$ and 
 modulus $k$ (see \mySecs{sec:2} and \ref{sec:4}).
 If the form can fit the correlation function data independently of $Q$, it provides strong numerical evidence for the wide applicability of \myEq{eq:3.5}, including unsolvable cases.
 Simultaneously, it leads us to the conjecture that triangular lattice models that satisfy the three conditions in \mySec{sec:2} can exhibit a unique ACL identical to the Ising model. 
 
 Because the fittings suffer from corrections that originate from the second-band of eigenvalues, annuli with a larger cut-off $c_{\rm max}$ than the Ising case should be used
 \cite{Fujimoto2020}.
 Following the same procedure as the Ising case, we optimize the cut-off as $c_{\rm max}=10^{-4},~10^{-4},\text{~and~}5\times10^{-5}$ for $Q=1$, 3, and 4, respectively; a finer optimization may be possible by taking temperature dependence into account, but we avoided it for clarity.

 We summarize the results in Table~III.
 Compared with the Ising case, the fitting conditions worsen because the correlation function data include larger errors in outer regions.  
 Additionally, the decrease of $|\myann|$ may cause a lowering of the statistical accuracy of the estimates of optimal values.
 Despite this, we find that our procedure estimates $\bar A$ and $\bar k$ within four or five-digit accuracy based on the following observations: 
 First, deep in the disordered phase, we theoretically expect the amplitude to be $A\simeq 1$. 
 The estimates $\bar A$ in the table agree with this condition and converge to 1 for large $t$ independently of $Q$.
 Second, in the second row of \myFig{fig:colorMap}, we provide color maps of reduced residual errors in fittings between $\bar{\cal F}_{\rm tri}(i,j)$ and $c(i,j)$ defined by
 \begin{equation}
 {\cal R}_{\rm tri}(i,j)=\frac{\bar{\cal F}_{\rm tri}(i,j)-c(i,j)}{c(i,j)}.
 \label{eq:3.7}
 \end{equation}
 One hexagon corresponds to each site $(i,j)$, whose color represents the absolute value $|{\cal R}_{\rm tri}|$ and whose boundary line represents its sign, that is, we draw boundary lines for hexagons if the residual errors are positive.
 We see that the optimized form asymptotically fits the MC data in all directions, excluding the central circular domain.
 Compared with the  $Q=2$ case, the directional dependence of residuals is visible for $Q\ne2$, which can be attributed to the second-band eigenvalue corrections; see \mySec{sec:2.B}. 
 As a result, we confirm that the form (\ref{eq:3.5}) can also fit unsolvable models' asymptotic correlation functions, although their accuracy is lower by about 1 or 2 digits than that in the Ising case.

 \subsection{Comparison with the Ornstein--Zernike form}
 \label{sec:3.C}
 
 Following the square lattice case 
 \cite{Fujimoto2020}, 
 we provide a second example in which the form extracted from the three conditions (i)--(iii) combined with the lattice symmetry (see \mySec{sec:2}) well describes the asymptotic behavior of correlation functions.
 Indeed, we observed that \myEq{eq:3.5} could fit the correlation functions of the triangular lattice $Q$-state Potts model and clarified its wide applicability.

 Now we compare fitting qualities between the present form and OZ form: ${\cal F}_{\rm OZ}(i,j;B,\xi)=B e^{-R/\xi}/\sqrt{R}~(R\ne0)$. 
 ${\cal F}_{\rm OZ}$ has been widely used to analyze the correlation functions in disordered phases; however, it possesses continuous rotational symmetry. 
 Therefore, it does not clarify the discreteness effects that we focus on in this study.

 To clarify the difference in degree consistent with MC data, we perform the same fitting calculations using $\chi^2_{\rm OZ}(B,\xi)$ statistics, that is, we replace the measure ${\cal F}_{\rm tri}$ in \myEq{eq:3.6} with ${\cal F}_{\rm OZ}$, but keep the annular regions $\cal D$ the same in both cases.
 We denote the optimized value by
 $\bar B$ and $\bar\xi$,
 and define the reduced residual errors of fittings as 
 ${\cal R}_{\rm OZ}(i,j)=[\bar{\cal F}_{\rm OZ}(i,j)-c(i,j)]/c(i,j)$,
 where
 $\bar{\cal F}_{\rm OZ}(i,j)={\cal F}_{\rm OZ}(i,j;\bar B,\bar\xi)$. 
 The color maps of the residuals are displayed in the first row of \myFig{fig:colorMap}. 
 
 We find that, although the number of free parameters is the same, the quality of fitting using the OZ form is much lower than that using the form (\ref{eq:3.5}). 
 This discrepancy indicates the existence of large systematic errors in the OZ form and identifies its insufficiency in terms of describing the off-critical correlation functions.
 As the comparison in the $Q=2$ case demonstrates, although ${\cal R}_{\rm tri}$ is seemingly rotational symmetric, ${\cal R}_{\rm OZ}$ is $C_{6v}$ symmetric in direction, which reflects the lack of discrete properties in the OZ form.
 Similarly, comparing other cases demonstrates the advantage of the present $C_{6v}$ form. 
 \FIGcolorMap
 Meanwhile, ${\cal R}_{\rm tri}$ clearly shows an oscillation accompanied by sign changes in the angular direction.
 Intriguingly, ${\cal R}_{\rm tri}<0$ and ${\cal R}_{\rm tri}>0$ in the row and diagonal directions, respectively for $Q=1$, whereas they take the opposite sign for $Q=3$ and 4.
 Because the correction from the second-band eigenvalues mainly contributes to the residual errors, its sign for $Q<2$ is seemingly the opposite of that for $Q>2$.
 Indeed, this prediction is consistent with the exact result (and the numerical result) that the correction from the second-band eigenvalues \myEq{eq:2.9}  vanishes for $Q=2$.
 \newpage
 \section{DISCUSSION AND SUMMARY}
 \label{sec:4}
 We investigated asymptotic correlation functions of the $Q$-state Potts model on a triangular lattice. 
 In \mySec{sec:2}, following the argument for $C_{4v}$ \cite{Fujimoto2020,Fujimoto2002}, we constructed asymptotic forms for $C_{6v}$. 
 First, we took the three properties (i)--(iii) into account, which are directly connected with $C_{2v}$ symmetry.
 Our exact analyses showed that the asymptotic correlation function of the triangular lattice Ising model satisfies (i)--(iii).
 We expect that the three properties are widely applicable to the models on the triangular lattice, whether solvable or not. 
 Then, we found that the asymptotic correlation function is written as integrals of the products of sn functions.
 Based on this integral representation, we derived the asymptotic form for $C_{6v}$ using the fact that $C_{2v}$ is the normal subgroup of $C_{6v}$.
 The product structure of the sn functions is essentially the same as that of $C_{4v}$ \cite{Fujimoto2020}.
 By contrast, unlike the $C_{4v}$ case, where the fitting forms include three or more free parameters, we found the minimal case, which has only two parameters: the normalization constant $A$ and modulus $k$.

 From the exact analyses for $Q=2$, we indicated that the minimal case applies to the general-$Q$ Potts model above transition temperatures.
 We performed MC simulations for $Q=1,~2,~3,~4$ above $\myTc(Q)$ and then successfully fit the MC data about five-digit accuracy.
 It is worth noting that, although there were fewer free parameters, we performed the fittings with the same accuracy as the square-lattice model calculations \cite{Fujimoto2020}.
 The present observation indicates the validity of the minimal form for correlation functions and the efficiency of using our approach to study triangular lattice models. 

 In the following, to clarify the physical meaning of the minimal form, we discuss the ECS \cite{Wulff1901,Burton1951}.
 We show that the ECS derived from the ACL is given by a simple algebraic curve of genus 1. 
 Furthermore, the product structures of sn functions relate to differential forms on the algebraic curve.
 Using birational transformations, we indicate an important role of modulus $k$ in representing the weak universality concept \cite{Suzuki1974} in critical phenomena.

 \subsection{Equilibrium crystal shape for the honeycomb lattice}
 \label{sec:4.A}

 In this subsection, we derive the ECS for the honeycomb lattice from the asymptotic correlation function in \mySec{sec:3}. 
 Suppose that $i$ and $j$ become large with $i/j$ fixed to be a constant in \myEq{eq:3.5}. 
 We introduce angle $\theta$ between the directions of ${\bf a}_1$ and $j{\bf a}_1+i{\bf a}_2$ as follows:
 \begin{equation}
 R\cos\theta
 =j-
 \frac{1}{2}i,
 \quad
 R\sin\theta
 =
 \frac{\sqrt 3}{2}i
 \quad
 {\rm with~}
 R=\sqrt{j^2+i^2-ij}
 \label{eq:4.1}
 \end{equation}
 (see \myFig{fig:1}).  
 We estimate the integral on the right-hand side using the method of steepest descent. 
 We calculate ACL $\xi$ as follows:
 \begin{align}
 -\frac{1}{\xi}
 = 
 \frac{2}{\sqrt 3}\bigg\{
 \cos(\theta-\frac{\pi}{6})
 &\ln\mySnSn{\phi_{\rm s}}{+\frac{\myIm I'}{6}}{-\frac{\myIm I'}{6}}
 \cr
 +
 \sin\theta
 &\ln\mySnSn{\phi_{\rm s}}{-\frac{\myIm I'}{2}}{-\frac{5\myIm I'}{6}}
 \bigg\},
 \label{eq:4.2}
 \end{align}
 where we determine the saddle point $\phi_{\rm s}$ as a function of $\theta$ by
 \begin{align}
 &
  \cos(\theta-\frac{\pi}{6})~
  \mySn{2\phi_{\rm s}}
 \sinh\left\{\ln\mySnSn{\phi_{\rm s}}{+\frac{\myIm I'}{6}}{-\frac{\myIm I'}{6}}\right\}
 \cr 
 &+
 \sin\theta~
 \mySn{2\phi_{\rm s}-\frac{4\myIm I'}{3}}
 \sinh\left\{\ln\mySnSn{\phi_{\rm s}}{-\frac{\myIm I'}{2}}{-\frac{5\myIm I'}{6}}\right\}
 =0
 \label{eq:4.3}
 \end{align}
 with the condition $\phi_{\rm s}=I$ for $\theta=0$.
 Then, the duality \cite{Fujimoto1999} relates ACL $\xi$ on the triangular lattice to the anisotropic interfacial tension $\gamma^{\ast}_{\rm h}$ below the transition temperature on the honeycomb lattice as follows:
 \begin{equation}
 \frac{\gamma^{\ast}_{\rm h}}{\mykB T^{\ast}_{\rm h}}=\frac{1}{\xi}
 \text{~~in~all~directions}
 \label{eq:4.4}
 \end{equation}
 (also see \myRefs{Holzer1990a} and \cite{Zia1986}).
 The reduced interaction constant on the honeycomb lattice $J_{\rm h}/\mykB T^{\ast}_{\rm h}$ is given by
 \begin{equation}
 {\rm e}^{2J_{\rm h}/\mykB T^{\ast}_{\rm h}}-1
 =
 \frac{Q}{{\rm e}^{2J/\mykB T}-1}.
 \label{eq:4.5}
 \end{equation}

 From $\gamma^{\ast}_{\rm h}$, we can determine the ECS for the honeycomb 
 lattice using Wulff's construction: 
 We denote the point on the ECS by $(X,Y)$, use \exEq{(4.2)} in \myRef{Fujimoto2020} with $\theta_{\perp}$ replaced by $\theta$, and define the exponentials as follows:
 \begin{equation}
 \alpha
 =
 \exp\left(-\Lambda X\right),
 \quad
 \beta
 =
 \exp\left[-\Lambda\left(\!\frac{\sqrt{3}Y}2-\frac{X}2\right)\right],
 \label{eq:4.6}
 \end{equation}
 where $\Lambda$ is a scale factor used to adjust an area of the ECS.
 Then, the ECS is given by
 \begin{align}
 \alpha
 =
 \mysnsn{\phi}{+\frac{\myIm I'}{6}}{-\frac{\myIm I'}{6}},
 \quad
 \beta
 =
 \mysnsn{\phi}{-\frac{\myIm I'}{2}}{-\frac{5\myIm I'}{6}}.
 \label{eq:4.7}
 \end{align}
 As $\phi$ moves from $I$ to $I+2\myIm I'$ on the line $\Re(\phi)=I$, for example, $(X,Y)$ sweeps out the ECS; see \exFig{5} in \myRef{Zia1986} or \exFig{6(a)} in \myRef{Fujimoto2002a}. 
 Generally, the ECSs are written in a compact form \cite{Holzer1990}.
 In the present case, we can rewrite \myEq{eq:4.7} as 
 \begin{equation}
 \alpha^2\beta^2+1+(\alpha\beta+1)(\alpha+\beta)+H\alpha\beta
 =
 0
 \label{eq:4.8}
 \end{equation}
 with
 \begin{equation}
 H
 =
 2\frac{\mydn^3(\frac{2\myIm I'}{3})+1}{k^2\mysn^2(\frac{2\myIm I'}{3})}. 
 \label{eq:4.9}
 \end{equation}

 \subsection{Birational transformations among algebraic curves}
 \label{sec:4.B}
 
 Equation~(\ref{eq:4.8}) defines an algebraic curve on the $\alpha\beta$-plane. 
 By introducing the homogeneous coordinate, we find two nodes at its infinity \cite{Namba1984,Walker1950}. 
 According to the so-called genus formula in \exChap{2.1} of \myRef{Namba1984}, \myEq{eq:4.8} is an algebraic curve of genus 1. 
 With this in mind, we return to the correlation functions in the minimal case.
 Using \myEq{eq:4.8}, we can re-express \myEq{eq:3.5} as
 \begin{equation}
  {\cal F}_{\rm tri}(i,j;A,k)
  =
  A\oint {d\alpha}
   \oint {d\beta}
   \frac{\alpha^j\beta^i}{\alpha^2\beta^2+1+(\alpha\beta+1)(\alpha+\beta)+H\alpha\beta},
 \label{eq:4.10}
 \end{equation}
 where the contour integrals are performed along the unit circles on the complex planes 
 \cite{Vaidya1976,Holzer1990a,Holzer1990,Akutsu1990}. 
 This expression shows that the product structure of sn functions in \myEq{eq:2.8} can be regarded as polynomials on the algebraic curve~(\ref{eq:4.8}) on the $\alpha\beta$-plane.

 Although essentially the same expression as \myEq{eq:4.10} was found in the exact calculation for $Q=2$ at $T>\myTc$ \cite{Vaidya1976}, our analysis demonstrates that the product structure on the algebraic curve~(\ref{eq:4.8}) is a direct consequence of $C_{6v}$ symmetry, which is independent of solvability and thus universal.

 As mentioned in \mySec{sec:2}, for $T<\myTc(Q)$, we expect the leading asymptotic behavior of the correlation function to be given by \myEq{eq:2.9}, that is, the form in the next to minimal case.
 According to the analysis in \mySec{sec:4.A}, to find the minimal case, we should consider the anisotropic interfacial tension that is related by duality to the ACL of the honeycomb lattice Potts model in the disordered phase. 
 Thus, a numerical study of the honeycomb lattice Potts model is important and is now in progress (in this respect, note that the minimal case was found in the antiferroelectric ordered regime of the Kagome lattice eight-vertex model \cite{Fujimoto2002a}).

 Now, we consider the role of the algebraic curve~(\ref{eq:4.8}) as a universal scale that measures the amount of deviation from criticalities.
 As observed, it is common among models with $C_{6v}$ symmetry and contains only one parameter $H$, or equivalently, $k$ by which we determine $H$ via \myEq{eq:4.9}.
 Additionally, it is independent of the types of criticalities.
 Hence, the shapes of the algebraic curve can specify the deviations from criticalities.
 Meanwhile, this argument is restricted to models with $C_{6v}$ symmetry.
 We can extend it to include a wider class of models with different lattice symmetries, for example, $C_{2v}$ and $C_{4v}$. 
 As mentioned in \myRef{Fujimoto2020}, in performing this extension, birational equivalence among algebraic curves plays a crucial role. 
 We demonstrate such an extension by considering $C_{4v}$ symmetry as an example.

 We obtained the asymptotic form for the square lattice $Q$-state Potts model, that is, ${\cal F}_{\rm sq}(i,j;A,k,b)$ \cite{Fujimoto2020}. 
 We denote the point on the ECS by $(X',Y')$, and write the exponentials as
 \begin{equation}
 \alpha'
 =
 \exp\left(-\Lambda X'\right),
 \quad
 \beta'
 =
 \exp\left(-\Lambda Y'\right).
 \label{eq:4.7b}
 \end{equation}
 We find that they satisfy \myEq{eq:1.2}, that is,
 \begin{equation}
 \alpha'^2\beta'^2
 +1
 +{\bar A}_2(\alpha'\beta'+1)(\alpha'+\beta')
 +\alpha'^2+\beta'^2
 +{\bar A}_4\alpha'\beta'
 =
 0.
 \label{eq:4.12}
 \end{equation}
 This is also an algebraic curve of genus 1.
 If the values of $k$ in \myEqs{eq:4.8} and (\ref{eq:4.12}) are the same, we can suitably choose rational functions to connect the variables of these two curves:
 \begin{equation}
 \alpha'
 =
 \frac{\alpha+c}{1+c\alpha},
 \quad
 \beta'
 =
 \frac{\Phi_1(\alpha,\beta)}{\Phi_2(\alpha,\beta)}
 \label{eq:4.13}
 \end{equation}
 with
 \begin{equation}
 c
 =
 -\mysnsn{b\frac{\myIm I'}{4}}{+\frac{\myIm I'}{6}}{-\frac{\myIm I'}{6}},
 \label{eq:4.14}
 \end{equation}
 where $\Phi_1(\alpha,\beta)$ and $\Phi_2(\alpha,\beta)$ are the polynomials of $\alpha$, $\beta$.
 Because a lengthy calculation is required, we provide the details of their derivation in \myAppendix{appendix:B}.
 Additionally, we find the inverse transformation from $\alpha'$, $\beta'$ to $\alpha$, $\beta$; hence, the transformation is bidirectional.

 To understand the implications of the transformation, 
 we consider the critical limit by taking $k\rightarrow 1$.
 Using the conjugate modulus transformation, from \myEq{eq:4.13}, we obtain a transformation between the ECS in \mySec{sec:4.A} and that in \exSec{IV} of \myRef{Fujimoto2020}.
 It follows that
 \begin{equation}
 k\sim 1-8q',
 \qquad 
 q'=\exp\left(\!-\pi\frac{I}{I'}\right).
 \label{eq:4.15}
 \end{equation}
 Note that the conjugate nome $q'\sim 1/\xi$.  
 To fix the areas of the ECSs, we adjust the scale factor as follows:
 \begin{equation}
 \Lambda=q'.
 \label{eq:4.16}
 \end{equation}
 Then, \myEqs{eq:4.6} and (\ref{eq:4.7b}) reduce to
 \begin{equation}
 \alpha
 \sim
 1-\Lambda X,
 \quad
 \beta
 \sim 
 1-\Lambda\left(\!\frac{\sqrt 3Y}{2}-\frac{X}{2}\right)
 \end{equation}
 and
 \begin{equation}
 \alpha'
 \sim 
 1-\Lambda X',
  \quad
 \beta'
 \sim 
 1-\Lambda Y',
 \label{eq:4.17}
 \end{equation}
 respectively.
 Equations (\ref{eq:4.8}) and (\ref{eq:4.12}) show that the ECSs become circles near the critical point; only their radii are different.
 From the transformation~(\ref{eq:4.13}), we find that   
 \begin{equation}
 \sqrt{X'^2+Y'^2}
 =
 \frac{\cos\frac{b\pi}{4}}{\cos\frac{\pi}{6}}
 \sqrt{X^2+Y^2}.
 \label{eq:4.18}
 \end{equation}
 
 To explain the continuously varying  exponents in the eight-vertex model \cite{Baxter2007,Johnson1973}, 
 Suzuki proposed the weak universality concept \cite{Suzuki1974}, where the inverse correlation length $1/\xi$ was regarded as a natural variable that measures the departure from critical points. 
 The relation (\ref{eq:4.18}) indicates that, to match the correlation length given in \mySec{sec:4.A} with that in \myRef{Fujimoto2020}, dilatation by the amount of $\cos({b\pi}/{4})/\cos({\pi}/{6})$ is required, which depends on both the degrees of freedom (like $Q$) and the types of lattices (e.g., triangle and square).
 In this respect, birational equivalence states that $k$ is more fundamental than $1/\xi$.

 Because the birational transformation~(\ref{eq:4.13}) is well-defined for general $k$, including the $k\to1$ limit, the modulus can provide a universal measure of the departure from critical points among various models defined on different lattices. 
 Mathematically, birational equivalence is a basic concept in the field of algebraic geometry \cite{Walker1950}. 
 In addition to the genus, $k$ is known as a birational invariant. 
 It is strongly suggested that the rich structures of birational geometry introduce a new insight into the study of lattice models. 
 We will report on our further investigations regarding this possibility in the future. 

 \acknowledgements
 We thank Professors
 Macoto Kikuchi and Yutaka Okabe
 for stimulating discussions again.
 The main computations were performed using the facilities in Tohoku University and Tokyo Metropolitan University.
 This research was supported by a grant-in-aid from KAKENHI No.~26400399.

 \appendix
 \makeatletter
 \renewcommand{\theequation}{\Alph{section}\arabic{equation}}
 \@addtoreset{equation}{section}
 \makeatother

 \newpage 
 \section{ASYMPTOTIC CORRELATION FUNCTIONS FOR $\boldsymbol{Q=2}$}
 \label{appendix:A}
 In \exApp{A} of \myRef{Fujimoto2020}, we investigated the correlation length in the square lattice Ising model 
 by extending the method of commuting transfer matrices in \exChap{7} of \myRef{Baxter2007} using the shift operator.
 In this appendix, we apply the same method to the Ising model on the triangular lattice. 
 We define inhomogeneous systems on the square lattice. Each system still possesses a one-parameter family of commuting transfer matrices.
 The products of commuting transfer matrices yield transfer matrices on the triangular lattice. 
 We analyze the asymptotic correlation function along 12 directions 
 and then find the three properties (i)--(iii) in \myRef{Fujimoto2020} that hold for the triangular lattice.
 This result supports our discussion on obtaining the asymptotic correlation functions with $C_{6v}$ symmetry in \mySec{sec:2}.

 \subsection{Inhomogeneous transfer matrices}
 \label{appendix:A1}

 First, we draw a square lattice diagonally. 
 For the Ising model, each $\sigma_{\bf r}$ takes the values of $\pm1$ 
 because it is defined as $\sigma_{\bf r}=\exp(\myIm\pi q_{\bf r})$ with $q_{\bf r}=0,1$ (see \mySec{sec:3}). 
 The Hamiltonian is written as \exEq{(A1)} in \myRef{Fujimoto2020}, 
 where the nearest-neighbor spins are coupled by $J$ or $J'$ depending on the direction. 
 Using Jacobi's elliptic functions, we parameterize the reduced coupling constants $K=J/\mykB T$, $K'=J'/\mykB T$ 
 using \exEq{(A5)} in \myRef{Fujimoto2020} for $T>\myTc$ and \exEq{(A6)} in \myRef{Fujimoto2020} for $T<\myTc$.
 To investigate the triangular lattice Ising model, we suppose that 
 the spectral parameter $u$ varies from site to site \cite{Baxter2007}. 
 We introduce a real number $v_0$ and define inhomogeneous transfer matrices. 
 We consider two successive rows, and let
 $\sigma =\{\sigma_0 ,\dots,\sigma_{N-1} \}$ and 
 $\sigma'=\{\sigma_0',\dots,\sigma_{N-1}'\}$ 
 be two sets of spins in the lower and upper rows, respectively ($N$ even); see \myFig{fig:figA1}.
 We assume periodic boundary conditions in both directions.
 Then, the transfer matrices ${\bf V}$ and ${\bf W}$ are given by elements as follows:
 \begin{align}
  [{\bf V}(u)]_{\sigma,\sigma'}
  &= 
  \exp
  \left\{
  \sum_{l=0}^{N-1}[K'(u)\sigma_l\sigma_l'+K(u+v_0)\sigma_{l+1}\sigma_l']
  \right\},
  \cr
  [{\bf W}(u)]_{\sigma,\sigma'}
  &= 
  \exp
  \left\{
  \sum_{l=0}^{N-1}[K(u)\sigma_l\sigma_l'+K'(u+v_0)\sigma_l\sigma_{l+1}']
  \right\},
 \label{eq:A1}
 \end{align}
 where
 $\sigma_N=\sigma_0$ and $\sigma_N'=\sigma_0'$.
 We regard modulus $k$ and $v_0$ as fixed constants. 
 They satisfy the following commutation relations:
 \begin{equation}
 [{\bf V}(u),{\bf V}(u')]=
 [{\bf W}(u),{\bf W}(u')]=
 [{\bf V}(u),{\bf W}(u')]=0
 \quad
 \forall u,~u'\in\mathbb{C}.
 \label{eq:A2}
 \end{equation}

 \FIGAone
 
 We denote the eigenvalues of ${\bf V}(u)$ and ${\bf W}(u)$ by $V(u)$ and $W(u)$, respectively.
 Then, as $N\rightarrow\infty$
 \begin{equation}
 V(u)\sim\kappa(u)^{\frac{N}2}\kappa(u+v_0)^{\frac{N}2},
 \quad 
 W(u)\sim\kappa(u)^{\frac{N}2}\kappa(u+v_0)^{\frac{N}2},
 \label{eq:A3}
 \end{equation}
 where $\kappa(u)$ is given by \exEqs{(A14)--(A16)} in \myRef{Fujimoto2020}; 
 also see \exChap{11} of \myRef{Baxter2007}. 
 To calculate the asymptotic correlation function, we define the following limiting functions:
 \begin{equation}
 \lim_{N\rightarrow\infty}\frac{W(u)}{[\kappa(u)\kappa(u+v_0)]^{\frac{N}2}},
 \quad
 \lim_{N\rightarrow\infty}\frac{V(u)}{[\kappa(u)\kappa(u+v_0)]^{\frac{N}2}}.
 \label{eq:A4}
 \end{equation}
 In \myRef{Fujimoto2020}, we proved that they are the same form and satisfy \exEq{(A27)} or \exEq{(A28)}.
 Thus, we write both of them as
 \begin{equation}
 \pm\prod_{l=1}^{\mu}k^{\frac12}
 \mySn{-\phi_l+\frac{\myIm I'}{2}+\myIm u+\frac{\myIm v_0}{2}}. 
 \label{eq:A5}
 \end{equation}
 As a result, we label the limiting function using $\mu$-real variables $\phi_1,\dots,\phi_\mu$. 
 We denote it by $L(\phi_1,\dots,\phi_\mu|u)$. 
  
 To investigate the triangular lattice Ising model with the isotropic interaction, 
 we set $v_0=I'/3$ and take the limits of $u$ as follows:
 The transfer matrix can be constructed as
 \begin{equation}
 {\bf Y}
 =
 \lim_{\substack{u_1\rightarrow0\\u_2\rightarrow 2v_0}}
 {\bf W}\left(\frac{I'}{3}\right)
 \frac{{\bf V}(u_1)}{\kappa(u_1)^{\frac{N}{2}}}
 {\bf W}\left(\frac{I'}{3}\right)
 \frac{{\bf V}(u_2)}{\kappa(u_2+v_0)^{\frac{N}{2}}}, 
 \label{eq:A6}
 \end{equation}
 and the shift operator as
 \begin{equation}
 {\bf X}
 =
 \lim_{\substack{u_1\rightarrow I'\\u_2\rightarrow 2v_0}}
 \frac{{\bf W}(u_1)}{[\kappa(u_1)\kappa(u_1+v_0)]^{\frac{N}{2}}}
 \frac{{\bf V}(u_2)}{[\kappa(u_2)\kappa(u_2+v_0)]^{\frac{N}{2}}}.
 \label{eq:A7}
 \end{equation}
 We showed in \exApp{A} of \myRef{Fujimoto2020} that the asymptotic correlation function is calculated from two ratios $L_Y$ and $L_X$; the former (latter) denotes the ratio between the eigenvalues and the largest eigenvalue of {\bf Y} ({\bf X}).
 Using $L(\phi|u)$, they are represented as 
 \begin{equation}
 L_X(\phi)=L(\phi|I')L\left(\!\phi\Big|\frac{2I'}{3}\right),
 \quad 
 L_Y(\phi)=L\left(\!\phi\Big|\frac{I'}{3}\right)L(\phi|0)L\left(\!\phi\Big|\frac{I'}{3}\right)
 L\left(\!\phi\Big|\frac{2I'}{3}\right).
 \end{equation}

 In the $N\rightarrow\infty$ limit, the summation over eigenvalues becomes integrals because of their continuous distribution.
 In particular, for $T>\myTc$, we calculate the leading asymptotic behavior of the correlation function from a band of the next-largest eigenvalues with $\mu=1$. 
 For ${\bf r}=j{\bf a}_1+i{\bf a}_2$ (see \myFig{fig:1}), it is given by the integral with respect to $\phi_1$ as
 \begin{align}
&\langle\sigma_{\bf o}\sigma_{\bf r}\rangle
 \sim
 {\rm const}\int^I_{-I}\ d\phi_1
 \cr
&\mySnSn{-\phi_1}{-\frac{\myIm I'}{3}}{-\frac{2\myIm I'}{3}}^j
 \mySnSn{-\phi_1}{-\myIm I'}{-\frac{4\myIm I'}{3}}^i
 \label{eq:A8}
 \end{align}
 with
 \begin{equation}
 \sinh K=\frac{\myIm}{\mySn{\frac{2\myIm I'}{3}}}.
 \label{eq:A8a}
 \end{equation}
 Because the normalization constant is identical to that of the square lattice Ising model, 
 \begin{equation}
 {\rm const}=\frac{(1-k^2)^{\frac14}}{\pi}
 \label{eq:A8b}
 \end{equation}
 \cite{Baxter1978}; see also \exChap{11} of \cite{Baxter2007}. 
 Note that the minimal form~(\ref{eq:2.8}) with $(\nu,\nu')=(2,4)$ coincides with \myEq{eq:A8} if we change the integration variable to $\Theta=-\phi_1-\myIm I'/2$ 
 and shift the integration path suitably. 
 Because of $\mathbb{Z}_2$ symmetry, the contribution 
 of the next-to-next-largest eigenvalues with $\mu=2$ vanishes.
 Therefore, the first correction to the asymptotic behavior (\ref{eq:A8}) originates from $\mu=3$ (see the numerical results in \mySec{sec:3.C}).

 For $T<\myTc$, the two largest eigenvalues are asymptotically degenerate as $N\rightarrow\infty$.
 The next-largest eigenvalues correspond to the case $\mu=2$.
 Thus, the asymptotic correlation function is given by the double integral with respect to $\phi_1,\phi_2$ as
 \begin{align}
 &\langle\sigma_{\bf o}\sigma_{\bf r}\rangle
 -
 \langle\sigma_{\bf o}\rangle\langle\sigma_{\bf r}\rangle
 \sim 
 \int^I_{-I}\ d\phi_1\int^I_{-I}\ d\phi_2~\rho(\phi_1,\phi_2) 
 \cr
 \times&
 \mySnSn{-\phi_1}{-\frac{\myIm I'}{3}}{-\frac{2\myIm I'}{3}}^j
 \mySnSn{-\phi_1}{-\myIm I'}{-\frac{4\myIm I'}{3}}^i
 \cr
 \times&
 \mySnSn{-\phi_2}{-\frac{\myIm I'}{3}}{-\frac{2\myIm I'}{3}}^j
 \mySnSn{-\phi_2}{-\myIm I'}{-\frac{4\myIm I'}{3}}^i,
 \label{eq:A9}
 \end{align}
 with
 \begin{equation}
 \sinh K=\frac{\myIm}{k\mySn{\frac{2\myIm I'}{3}}},
 \label{eq:A9a}
 \end{equation}
 where we determine the function $\rho(\phi_1,\phi_2)$ from the distribution of the eigenvalues and the matrix elements \cite{Johnson1973}.
 Note that the next to minimal case with $(\nu,\nu')=(4,8)$, 
that is, \myEq{eq:2.9} reproduces \myEq{eq:A9} by a suitable transformation.

 \subsection{Passive rotations}
 \label{appendix:A2}
 
 We consider transfer matrices along various directions to find the role of $C_{6v}$ symmetry. 
 In \exApp{A} of \myRef{Fujimoto2020}, we defined the Ising model on a square lattice rotated  
 through an arbitrary angle. 
 We found that the lattice rotations shift the integration paths with their deformations. 
 The same is expected to occur in the triangular lattice Ising model.  
 However, analyzing a triangular lattice rotated through an arbitrary angle is quite complicated. 
 Instead, we investigate transfer matrices along 12 directions to derive the three properties (i)--(iii). 
 
 We consider calculations of the ACLs. 
 For example, above $\myTc$ ($\mu=1$) we can estimate the integral in \myEq{eq:A8} using the method of steepest descent; see \mySec{sec:4.A}. 
 To calculate the correlation length along the horizontal direction $\theta=0$, we take the $j\rightarrow\infty$ limit with $i=0$ and then find the saddle point at $\phi_1=-\myIm I'/2+I$; for the definition of $\theta$, see \myEq{eq:4.1}. 
 When $\theta$ increases as $\theta=n\pi/6$ with $n=1,2,\dots,5$, we find that the saddle point moves on the line $\Re(\phi_1)=I$; it is located at $\phi_1=-\myIm I'/2-\myIm nI'/6+I$. 
 
 The increase of $\theta$ corresponds to active rotations. 
 Meanwhile, the method using passive rotations yields the same result. 
 First, we consider the triangular lattice rotated clockwise by angle $n\pi/3$ ($n=1,2,\dots,5$). 
 If we define the transfer matrix and shift operator on the rotated lattice, they are identical to $\bf Y$ and $\bf X$, respectively.  
 We repeat the analysis from \myEq{eq:A1} to \myEq{eq:A5}; the limiting function is given by \myEq{eq:A5} with $\phi_l$ replaced by $\bar{\phi_l}$ if we choose a suitable relation between them.
 
 Above $\myTc$, we rename $\phi_1$ on the right-hand side of \myEq{eq:A8} as $\bar{\phi}_1$. 
 Comparing the saddle point on the ${\bar\phi}_1$-plane with that on the $\phi_1$-plane, we find that $\phi_1$ and ${\bar \phi}_1$ 
 are related as ${\bar\phi}_1=\phi_1-\myIm nI'/3$. 
 Note that
 \begin{equation}
 L_X\left(\!\phi_1+\frac{\myIm I'}{3}\right)^2=L_X(\phi_1)L_Y(\phi_1),
 \quad 
 L_Y\left(\!\phi_1+\frac{\myIm I'}{3}\right)^2=L_X(\phi_1)^{-3}L_Y(\phi_1).
 \label{eq:A10}
 \end{equation}
 We extend these relations into the cases with $\mu>1$, which indicates that ${\bar \phi}_l=\phi_l-\myIm nI'/3$ for all $l$. 
 We find that the lattice rotation by $n\pi/3$ shifts the integration paths by $-\myIm n I'/3$ without deformations. 
 In fact, for $T<\myTc$, we start with \myEq{eq:A9} with $\phi_1,\phi_2$ replaced by $\bar{\phi}_1,\bar{\phi}_2$. 
 We obtain the same integration-path shifts caused by the $n\pi/3$-lattice rotations.
  
 Second, to investigate the case $(2n-1)\pi/6~(n=1,2,\dots,6)$, we define inhomogeneous transfer matrices as follows:  
 \begin{align}
  [\tilde{\bf V}(u)]_{\sigma,\sigma'}
  &= 
  \exp
  \Bigg\{
  \sum_{l=0}^{N-1}[
  \varepsilon_{2l,  2l  }^{(0)}+\epsilon_{2l+1,2l  }^{(1)}+
  \varepsilon_{2l+1,2l+1}^{(2)}+\epsilon_{2l+2,2l+1}^{(3)}
  ]
  \Bigg\},
  \cr
  [\tilde{\bf W}(u)]_{\sigma,\sigma'}
  &= 
  \exp
  \Bigg\{
  \sum_{l=0}^{N-1}[
  \epsilon_{2l,  2l  }^{(0)}+\varepsilon_{2l  ,2l+1}^{(1)}+
  \epsilon_{2l+1,2l+1}^{(2)}+\varepsilon_{2l+1,2l+2}^{(3)}
  ]
  \Bigg\},
 \label{eq:A11}
 \end{align}
 where
 $\sigma_{2N}=\sigma_0$ and $\sigma_{2N}'=\sigma_0'$. 
 We denote local energies between $\sigma_l$ and $\sigma'_{l'}$ with coupling constants
 $K(u)$, $K(u+v_0)$, $K(u+I'-v_0)$ and $K(u+v_0)$
 [$K'(u)$, $K'(u+v_0)$, $K'(u+I'-v_0)$ and $K'(u+v_0)$] 
 as
 $\epsilon_{l,l'}^{(0)}$, $\epsilon_{l,l'}^{(1)}$, $\epsilon_{l,l'}^{(2)}$ and $\epsilon_{l,l'}^{(3)}$
 [$\varepsilon_{l,l'}^{(0)}$, $\varepsilon_{l,l'}^{(1)}$, $\varepsilon_{l,l'}^{(2)}$ and $\varepsilon_{l,l'}^{(3)}$],
 respectively.
 Then, the transfer matrix in the rotated system is given by
 \begin{equation}
 \tilde{\bf Y}=\lim_{\substack{u_1\rightarrow v_0\\ u_2\rightarrow 0}}
 \frac{\tilde{\bf W}(u_1)}{\kappa(u_1+I'-v_0)^{\frac{N}{2}}}
 \frac{\tilde{\bf V}(u_2)}{\kappa(u_2)^{\frac{N}{2}}}. 
 \label{eq:A12}
 \end{equation}
 We cannot construct the shift operator $\tilde{\bf X}$ in a similar manner.
 Despite this, we calculate the correlation length along the direction $(2n-1)\pi/6$ from the limiting functions corresponding to $\tilde{\bf Y}$ using the method of steepest descent. 
 We find that the distances between the saddle points and integration paths change, which means that the lattice rotation by $(2n-1)\pi/6$ shifts the integration paths by $-\myIm (2n-1) I'/6$. 

 As a result, the lattice rotation clockwise by $n\pi/6$ causes the integration paths to shift by $-\myIm n I'/6~(n=1,2,\dots,11)$. 
 The ACL calculated on the rotated lattice must be the same as that on the original lattice. We derive the equivalence with the help of the analyticity of the limiting functions (or eigenvalues). 
 We find that (i) analyticity of the eigenvalues is required to ensure equivalence between the results in analyses along various directions. 
 As shown in \myEq{eq:A10}, we also find relations connected with the coordinate transformations for even $n$. 
 The product structures of the sn functions in \myEq{eq:A5} can derive the sixfold rotational symmetry. A necessary condition is that each of the limiting functions satisfies the inversion relation, that is, the first equation of \exEq{(A27)} or \exEq{(A28)} in \myRef{Fujimoto2020}.
 Thus, (ii) the limiting functions should satisfy the equation corresponding to $\pi$-rotational symmetry.
 Note that if we assume (iii) doubly periodic structures, we obtain \myEq{eq:A5}. 
 We expect that (iii) is generally satisfied for lattice systems; see \mySec{sec:2}. 

 \newpage
 \section{BIRATIONAL TRANSFORMATIONS}
 \label{appendix:B}
  In this appendix, we consider a relation between the algebraic curves (\ref{eq:4.8}) and (\ref{eq:4.12}); for the latter, see \CHK{\mySec{sec:4}} of \myRef{Fujimoto2020}. 
 The set of $\alpha$ and $\beta$ in \myEq{eq:4.8} is a basis of an elliptic function field. 
 Alternatively, the set of $\alpha$ and $\beta$ in \myEq{eq:4.12} is another example of a basis of the same elliptic function field. 
 If this is the case, the algebraic geometry indicates the possibility that rational functions relate to these two bases \cite{Walker1950}. 
 The relation is called a birational correspondence, and the two curves are considered to be birationally equivalent. 
 We clarify a condition necessary for this possibility.
 
 We start with the algebraic curve~(\ref{eq:4.12}) found in the $C_{4v}$ case:
 \begin{equation}
 \alpha'^2\beta'^2
 +1
 +{\bar A}_2(\alpha'\beta'+1)(\alpha'+\beta')
 +\alpha'^2+\beta'^2
 +{\bar A}_4\alpha'\beta'
 =
 0.
 \label{eq:B1}
 \end{equation}
 Using the sn functions, we parameterize \myEq{eq:B1} as follows:
 \begin{equation}
 \alpha'
 =
 \mysnsn{\phi}{+b\frac{\myIm I'}{4}}{-b\frac{\myIm I'}{4}},
 \quad 
 \beta'
 =
 \mysnsn{\phi}{+b\frac{\myIm I'}{4}-\frac{\myIm I'}{2}}{-b\frac{\myIm I'}{4}-\frac{\myIm I'}{2}},
 \label{eq:B2}
 \end{equation}
 with
 \begin{equation}
 {\bar A}_2=\frac{2\mycn(b\frac{\myIm I'}{2})\mydn(b\frac{\myIm I'}{2})}
 {1+k\mysn^2(b\frac{\myIm I'}{2})},
 \qquad 
 {\bar A}_4=4-4\frac{(k^{\frac12}+k^{-\frac12})^2}
 {1+k\mysn^2(b\frac{\myIm I'}{2})},
 \label{eq:B3}
 \end{equation}
 where $b$ is a free parameter introduced in \CHK{\mySec{sec:3}} of \myRef{Fujimoto2020}; also see the errata \cite{Fujimoto2020e}.
 
 Equations~(\ref{eq:B1}) and (\ref{eq:4.8}) are algebraic curves of genus 1.
 Furthermore, if both have the same modulus $k$, we can relate them using rational functions as follows:
 Using the additional formula for Jacobi's elliptic functions, we obtain the first relation of \myEq{eq:4.13}: 
 \begin{equation}
  \alpha'=\frac{\alpha+c}{1+c\alpha}, 
  \label{eq:B4}
 \end{equation}
 where
 \begin{equation}
 c=-kS_+S_-,
 \quad 
 S_{\pm}=\mySn{v\pm\eta}, 
 \label{eq:B5}
 \end{equation}
 with $v=b{\myIm I'}/{4}$ and $\eta={\myIm I'}/{6}$.

 To find the second relation of \myEq{eq:4.13}, we define the following quantities:
 \begin{align}
 X_0
 =&
 S_+^2+S_-^2-\mysn^2(\eta)-\frac{c^2}{k^2\mysn^2(\eta)},
 \cr
 X_1
 =&
 S_+^2+S_-^2-c^2\mysn^2(\eta)-\frac{1}{k^2\mysn^2(\eta)},
 \cr
 X_2
 =&
 \frac{1-c^2}{k}[\mycn(2v)\mydn(2\eta)+\mydn(2v)\mycn(2\eta)]
 +2c\left[\frac{\mycn~\mydn}{k\mysn}(\eta)\right]^2, 
 \cr
 X_3
 =&
 \frac{1-c^2}{k}[\mycn(2v)\mydn(2\eta)+\mydn(2v)\mycn(2\eta)]
 -2c\left[\frac{\mycn~\mydn}{k\mysn}(\eta)\right]^2, 
 \cr
 X_4
 =&
 4S_+^2+4S_-^2
 +2(1+c^2)
 \left[
 \frac{2\mycn^2(\eta)\mydn^2(\eta)-1}{k^2\mysn^2(\eta)}-\mysn^2(\eta)
 \right].
 \end{align}
 Then, we can prove that
 \begin{equation}
 X_0(\beta'^2\beta^2+1)+
 X_1(\beta'^2+\beta^2)+
 X_2(\beta'\beta^2+\beta')+
 X_3(\beta'^2\beta+\beta)+X_4\beta'\beta=0.
 \label{eq:B6}
 \end{equation}
 We multiply \myEq{eq:B1} by $(X_0\beta^2+X_1+X_2\beta)$, and \myEq{eq:B6} by $(\alpha'^2+{\bar A}_2\alpha'+1)$. 
 We subtract the latter from the former. 
 Then, it follows with the help of \myEq{eq:B4} that
 \begin{equation}
 \beta'=\frac{\Phi_1(\alpha,\beta)}{\Phi_2(\alpha,\beta)},
 \label{eq:B7}
 \end{equation}
 where $\Phi_1(\alpha, \beta)$ and $\Phi_2(\alpha, \beta)$ are the polynomials of $\alpha$ and $\beta$ given by
 \begin{align}
 \Phi_1(\alpha, \beta)
 =&
 (X_1-X_0)\{(c^2+{\bar A}_2c+1)(\alpha^2+1)+
 [{\bar A}_2(c^2+1)+4c]\alpha\}(\beta^2-1),  
 \cr
 \Phi_2(\alpha,\beta)
 =&
 \left\{[{\bar A}_2(c^2+1)+{\bar A}_4c](\alpha^2+1)+
 [{\bar A}_4(c^2+1)+4{\bar A}_2c]\alpha\right\}(X_0\beta^2+X_3\beta+X_1)  
 \cr
 &-
 \left\{(c^2+{\bar A}_2c+1)(\alpha^2+1)+[{\bar A}_2(c^2+1)+4c]\alpha\right\}
 (X_2\beta^2+X_4\beta+X_2).
 \label{eq:B8}
 \end{align}
 
 We derive the inverse transformation similarly: 
 From \myEq{eq:B4} we obtain
  \begin{equation}
  \alpha=\frac{\alpha'-c}{1-c\alpha'}.
  \label{eq:B9}
 \end{equation}
 We use \myEq{eq:4.8} instead of \myEq{eq:B1}.
 It follows that
 \begin{equation}
 \beta=\frac{\Psi_1(\alpha',\beta')}{\Psi_2(\alpha',\beta')},
 \label{eq:B10}
 \end{equation}
 where $\Psi_1(\alpha',\beta')$ and $\Psi_2(\alpha',\beta')$ are the fourth-order polynomials of $\alpha'$ and $\beta'$ given by
 \begin{align}
 \Psi_1(\alpha',\beta')
 =&
 (1-c)(\alpha'+1)\left[(c\alpha'-1)(X_0\beta'^2+X_2\beta'+X_1)+
 (\alpha'-c)(X_1\beta'^2+X_2\beta'+X_0)\right],
 \cr 
 \Psi_2(\alpha',\beta')
 =&
 \left\{(c^2-cH+1)(\alpha'^2+1)+[(c^2+1)H-4c]\alpha'\right\}(X_0\beta'^2+X_2\beta'+X_1)  
 \cr 
 &+
 (c-1)(\alpha'+1)(\alpha'-c)(X_3\beta'^2+X_4\beta'+X_3).
 \label{eq:B11}
 \end{align}
 Consequently, we obtain the birational transformation that connects the algebraic curves (\ref{eq:4.8}) and (\ref{eq:4.12}) \cite{Walker1950}.

 \newpage

\begin{thebibliography}{61}%
\makeatletter
\providecommand \@ifxundefined [1]{%
 \@ifx{#1\undefined}
}%
\providecommand \@ifnum [1]{%
 \ifnum #1\expandafter \@firstoftwo
 \else \expandafter \@secondoftwo
 \fi
}%
\providecommand \@ifx [1]{%
 \ifx #1\expandafter \@firstoftwo
 \else \expandafter \@secondoftwo
 \fi
}%
\providecommand \natexlab [1]{#1}%
\providecommand \enquote  [1]{``#1''}%
\providecommand \bibnamefont  [1]{#1}%
\providecommand \bibfnamefont [1]{#1}%
\providecommand \citenamefont [1]{#1}%
\providecommand \href@noop [0]{\@secondoftwo}%
\providecommand \href [0]{\begingroup \@sanitize@url \@href}%
\providecommand \@href[1]{\@@startlink{#1}\@@href}%
\providecommand \@@href[1]{\endgroup#1\@@endlink}%
\providecommand \@sanitize@url [0]{\catcode `\\12\catcode `\$12\catcode
  `\&12\catcode `\#12\catcode `\^12\catcode `\_12\catcode `\%12\relax}%
\providecommand \@@startlink[1]{}%
\providecommand \@@endlink[0]{}%
\providecommand \url  [0]{\begingroup\@sanitize@url \@url }%
\providecommand \@url [1]{\endgroup\@href {#1}{\urlprefix }}%
\providecommand \urlprefix  [0]{URL }%
\providecommand \Eprint [0]{\href }%
\providecommand \doibase [0]{https://doi.org/}%
\providecommand \selectlanguage [0]{\@gobble}%
\providecommand \bibinfo  [0]{\@secondoftwo}%
\providecommand \bibfield  [0]{\@secondoftwo}%
\providecommand \translation [1]{[#1]}%
\providecommand \BibitemOpen [0]{}%
\providecommand \bibitemStop [0]{}%
\providecommand \bibitemNoStop [0]{.\EOS\space}%
\providecommand \EOS [0]{\spacefactor3000\relax}%
\providecommand \BibitemShut  [1]{\csname bibitem#1\endcsname}%
\let\auto@bib@innerbib\@empty
\bibitem [{\citenamefont {Wulff}(1901)}]{Wulff1901}%
  \BibitemOpen
  \bibfield  {author} {\bibinfo {author} {\bibfnamefont {G.}~\bibnamefont
  {Wulff}},\ }\href
  {https://doi.org/https://doi.org/10.1524/zkri.1901.34.1.449} {\bibfield
  {journal} {\bibinfo  {journal} {Z. Kristallogr. Cryst. Mater.}\ }\textbf
  {\bibinfo {volume} {34}},\ \bibinfo {pages} {449 } (\bibinfo {year}
  {1901})}\BibitemShut {NoStop}%
\bibitem [{\citenamefont {von Laue}(1944)}]{Laue1944}%
  \BibitemOpen
  \bibfield  {author} {\bibinfo {author} {\bibfnamefont {M.}~\bibnamefont {von
  Laue}},\ }\href@noop {} {\bibfield  {journal} {\bibinfo  {journal} {Z.
  Kristallogr.}\ }\textbf {\bibinfo {volume} {105}},\ \bibinfo {pages} {124}
  (\bibinfo {year} {1944})}\BibitemShut {NoStop}%
\bibitem [{\citenamefont {Herring}(1951)}]{Herring1951}%
  \BibitemOpen
  \bibfield  {author} {\bibinfo {author} {\bibfnamefont {C.}~\bibnamefont
  {Herring}},\ }\href {https://doi.org/10.1103/PhysRev.82.87} {\bibfield
  {journal} {\bibinfo  {journal} {Phys. Rev.}\ }\textbf {\bibinfo {volume}
  {82}},\ \bibinfo {pages} {87} (\bibinfo {year} {1951})}\BibitemShut {NoStop}%
\bibitem [{\citenamefont {Burton}\ \emph {et~al.}(1951)\citenamefont {Burton},
  \citenamefont {Cabrera},\ and\ \citenamefont {Frank}}]{Burton1951}%
  \BibitemOpen
  \bibfield  {author} {\bibinfo {author} {\bibfnamefont {W.~K.}\ \bibnamefont
  {Burton}}, \bibinfo {author} {\bibfnamefont {N.}~\bibnamefont {Cabrera}},\
  and\ \bibinfo {author} {\bibfnamefont {F.~C.}\ \bibnamefont {Frank}},\ }\href
  {https://doi.org/10.1098/rsta.1951.0006} {\bibfield  {journal} {\bibinfo
  {journal} {Philos. Trans. R. Soc. London, Ser. A}\ }\textbf {\bibinfo
  {volume} {243}},\ \bibinfo {pages} {299} (\bibinfo {year} {1951})}
  \BibitemShut {NoStop}%
\bibitem [{\citenamefont {Abraham}\ and\ \citenamefont
  {Reed}(1974)}]{Abraham1974}%
  \BibitemOpen
  \bibfield  {author} {\bibinfo {author} {\bibfnamefont {D.~B.}\ \bibnamefont
  {Abraham}}\ and\ \bibinfo {author} {\bibfnamefont {P.}~\bibnamefont {Reed}},\
  }\href {https://doi.org/10.1103/PhysRevLett.33.377} {\bibfield  {journal}
  {\bibinfo  {journal} {Phys. Rev. Lett.}\ }\textbf {\bibinfo {volume} {33}},\
  \bibinfo {pages} {377} (\bibinfo {year} {1974})}\BibitemShut {NoStop}%
\bibitem [{\citenamefont {Abraham}\ and\ \citenamefont
  {Reed}(1976)}]{Abraham1976}%
  \BibitemOpen
  \bibfield  {author} {\bibinfo {author} {\bibfnamefont {D.~B.}\ \bibnamefont
  {Abraham}}\ and\ \bibinfo {author} {\bibfnamefont {P.}~\bibnamefont {Reed}},\
  }\href@noop {} {\bibfield  {journal} {\bibinfo  {journal} {Commun. Math.
  Phys.}\ }\textbf {\bibinfo {volume} {49}},\ \bibinfo {pages} {35} (\bibinfo
  {year} {1976})}\BibitemShut {NoStop}%
\bibitem [{\citenamefont {van Beijeren}(1977)}]{Beijeren1977}%
  \BibitemOpen
  \bibfield  {author} {\bibinfo {author} {\bibfnamefont {H.}~\bibnamefont {van
  Beijeren}},\ }\href {https://doi.org/10.1103/PhysRevLett.38.993} {\bibfield
  {journal} {\bibinfo  {journal} {Phys. Rev. Lett.}\ }\textbf {\bibinfo
  {volume} {38}},\ \bibinfo {pages} {993} (\bibinfo {year} {1977})}\BibitemShut
  {NoStop}%
\bibitem [{\citenamefont {Jayaprakash}\ \emph {et~al.}(1983)\citenamefont
  {Jayaprakash}, \citenamefont {Saam},\ and\ \citenamefont
  {Teitel}}]{Jayaprakash1983}%
  \BibitemOpen
  \bibfield  {author} {\bibinfo {author} {\bibfnamefont {C.}~\bibnamefont
  {Jayaprakash}}, \bibinfo {author} {\bibfnamefont {W.~F.}\ \bibnamefont
  {Saam}},\ and\ \bibinfo {author} {\bibfnamefont {S.}~\bibnamefont {Teitel}},\
  }\href {https://doi.org/10.1103/PhysRevLett.50.2017} {\bibfield  {journal}
  {\bibinfo  {journal} {Phys. Rev. Lett.}\ }\textbf {\bibinfo {volume} {50}},\
  \bibinfo {pages} {2017} (\bibinfo {year} {1983})}\BibitemShut {NoStop}%
\bibitem [{\citenamefont {Rottman}\ and\ \citenamefont
  {Wortis}(1981)}]{Rottman1981}%
  \BibitemOpen
  \bibfield  {author} {\bibinfo {author} {\bibfnamefont {C.}~\bibnamefont
  {Rottman}}\ and\ \bibinfo {author} {\bibfnamefont {M.}~\bibnamefont
  {Wortis}},\ }\href {https://doi.org/10.1103/PhysRevB.24.6274} {\bibfield
  {journal} {\bibinfo  {journal} {Phys. Rev. B}\ }\textbf {\bibinfo {volume}
  {24}},\ \bibinfo {pages} {6274} (\bibinfo {year} {1981})}\BibitemShut
  {NoStop}%
\bibitem [{\citenamefont {Avron}\ \emph {et~al.}(1982)\citenamefont {Avron},
  \citenamefont {van Beijeren}, \citenamefont {Schulman},\ and\ \citenamefont
  {Zia}}]{Avron1982}%
  \BibitemOpen
  \bibfield  {author} {\bibinfo {author} {\bibfnamefont {J.~E.}\ \bibnamefont
  {Avron}}, \bibinfo {author} {\bibfnamefont {H.}~\bibnamefont {van Beijeren}},
  \bibinfo {author} {\bibfnamefont {L.~S.}\ \bibnamefont {Schulman}},\ and\
  \bibinfo {author} {\bibfnamefont {R.~K.~P.}\ \bibnamefont {Zia}},\ }\href
  {http://stacks.iop.org/0305-4470/15/i=2/a=006} {\bibfield  {journal}
  {\bibinfo  {journal} {J. Phys. A: Math. Gen.}\ }\textbf {\bibinfo {volume}
  {15}},\ \bibinfo {pages} {L81} (\bibinfo {year} {1982})}\BibitemShut
  {NoStop}%
\bibitem [{\citenamefont {Zia}\ and\ \citenamefont {Avron}(1982)}]{Zia1982}%
  \BibitemOpen
  \bibfield  {author} {\bibinfo {author} {\bibfnamefont {R.~K.~P.}\
  \bibnamefont {Zia}}\ and\ \bibinfo {author} {\bibfnamefont {J.~E.}\
  \bibnamefont {Avron}},\ }\href {https://doi.org/10.1103/PhysRevB.25.2042}
  {\bibfield  {journal} {\bibinfo  {journal} {Phys. Rev. B}\ }\textbf {\bibinfo
  {volume} {25}},\ \bibinfo {pages} {2042} (\bibinfo {year}
  {1982})}\BibitemShut {NoStop}%
\bibitem [{\citenamefont {Lieb}\ and\ \citenamefont {Wu}(1972)}]{Lieb1972}%
  \BibitemOpen
  \bibfield  {author} {\bibinfo {author} {\bibfnamefont {E.~M.}\ \bibnamefont
  {Lieb}}\ and\ \bibinfo {author} {\bibfnamefont {F.~Y.}\ \bibnamefont {Wu}},\
  }\bibinfo {title} {Two-dimensional ferroelectric models},\ in\ \href@noop {}
  {\emph {\bibinfo {booktitle} {Phase Transitions and Critical Phenomena}}},\
  Vol.~\bibinfo {volume} {1},\ \bibinfo {editor} {edited by\ \bibinfo {editor}
  {\bibnamefont {C.Domb}}\ and\ \bibinfo {editor} {\bibfnamefont
  {M.}~\bibnamefont {S.Green}}}\ (\bibinfo  {publisher} {Academic Press,
  London},\ \bibinfo {year} {1972})\ pp.\ \bibinfo {pages}
  {332--490}\BibitemShut {NoStop}%
\bibitem [{\citenamefont {Baxter}(2007)}]{Baxter2007}%
  \BibitemOpen
  \bibfield  {author} {\bibinfo {author} {\bibfnamefont {R.}~\bibnamefont
  {Baxter}},\ }\href {https://books.google.it/books?id=G3owDULfBuEC} {\emph
  {\bibinfo {title} {Exactly Solved Models in Statistical Mechanics}}},\ Dover
  books on physics\ (\bibinfo  {publisher} {Dover Publications},\ \bibinfo
  {year} {2007})\BibitemShut {NoStop}%
\bibitem [{\citenamefont {Landau}\ and\ \citenamefont
  {Lifshitz}(1980)}]{Landau1980}%
  \BibitemOpen
  \bibfield  {author} {\bibinfo {author} {\bibfnamefont {L.~D.}\ \bibnamefont
  {Landau}}\ and\ \bibinfo {author} {\bibfnamefont {E.~M.}\ \bibnamefont
  {Lifshitz}},\ }\href@noop {} {\emph {\bibinfo {title} {Statistical Physics,
  Part 1, 3rd Edition}}},\ Vol.~\bibinfo {volume} {5}\ (\bibinfo  {publisher}
  {Butterworth Heinemann},\ \bibinfo {year} {1980})\BibitemShut {NoStop}%
\bibitem [{\citenamefont {Andreev}(1981)}]{Andreev1981}%
  \BibitemOpen
  \bibfield  {author} {\bibinfo {author} {\bibfnamefont {A.}~\bibnamefont
  {Andreev}},\ }\href@noop {} {\bibfield  {journal} {\bibinfo  {journal} {Sov.
  Phys. JETP}\ }\textbf {\bibinfo {volume} {53}},\ \bibinfo {pages} {1063}
  (\bibinfo {year} {1981})}\BibitemShut {NoStop}%
\bibitem [{\citenamefont {Akutsu}\ and\ \citenamefont
  {Akutsu}(1990)}]{Akutsu1990}%
  \BibitemOpen
  \bibfield  {author} {\bibinfo {author} {\bibfnamefont {Y.}~\bibnamefont
  {Akutsu}}\ and\ \bibinfo {author} {\bibfnamefont {N.}~\bibnamefont
  {Akutsu}},\ }\href {https://doi.org/10.1103/PhysRevLett.64.1189} {\bibfield
  {journal} {\bibinfo  {journal} {Phys. Rev. Lett.}\ }\textbf {\bibinfo
  {volume} {64}},\ \bibinfo {pages} {1189} (\bibinfo {year}
  {1990})}\BibitemShut {NoStop}%
\bibitem [{\citenamefont {Selke}\ and\ \citenamefont
  {Pesch}(1982)}]{Selke1982a}%
  \BibitemOpen
  \bibfield  {author} {\bibinfo {author} {\bibfnamefont {W.}~\bibnamefont
  {Selke}}\ and\ \bibinfo {author} {\bibfnamefont {W.}~\bibnamefont {Pesch}},\
  }\href@noop {} {\bibfield  {journal} {\bibinfo  {journal} {Z. Phys. B}\
  }\textbf {\bibinfo {volume} {47}},\ \bibinfo {pages} {335} (\bibinfo {year}
  {1982})}\BibitemShut {NoStop}%
\bibitem [{\citenamefont {Fujimoto}(1992)}]{Fujimoto1992}%
  \BibitemOpen
  \bibfield  {author} {\bibinfo {author} {\bibfnamefont {M.}~\bibnamefont
  {Fujimoto}},\ }\href {https://doi.org/10.1007/BF01049029} {\bibfield
  {journal} {\bibinfo  {journal} {J. Stat. Phys.}\ }\textbf {\bibinfo {volume}
  {67}},\ \bibinfo {pages} {123} (\bibinfo {year} {1992})}\BibitemShut
  {NoStop}%
\bibitem [{\citenamefont {Fujimoto}(1993)}]{Fujimoto1993}%
  \BibitemOpen
  \bibfield  {author} {\bibinfo {author} {\bibfnamefont {M.}~\bibnamefont
  {Fujimoto}},\ }\href {http://stacks.iop.org/0305-4470/26/i=10/a=004}
  {\bibfield  {journal} {\bibinfo  {journal} {J. Phys. A: Math. Gen.}\ }\textbf
  {\bibinfo {volume} {26}},\ \bibinfo {pages} {2285} (\bibinfo {year}
  {1993})}\BibitemShut {NoStop}%
\bibitem [{\citenamefont {Fujimoto}(1997)}]{Fujimoto1997}%
  \BibitemOpen
  \bibfield  {author} {\bibinfo {author} {\bibfnamefont {M.}~\bibnamefont
  {Fujimoto}},\ }\href {http://stacks.iop.org/0305-4470/30/i=11/a=011}
  {\bibfield  {journal} {\bibinfo  {journal} {J. Phys. A: Math. Gen.}\ }\textbf
  {\bibinfo {volume} {30}},\ \bibinfo {pages} {3779} (\bibinfo {year}
  {1997})}\BibitemShut {NoStop}%
\bibitem [{\citenamefont {Akutsu}\ and\ \citenamefont
  {Akutsu}(1987{\natexlab{a}})}]{Akutsu1987a}%
  \BibitemOpen
  \bibfield  {author} {\bibinfo {author} {\bibfnamefont {N.}~\bibnamefont
  {Akutsu}}\ and\ \bibinfo {author} {\bibfnamefont {Y.}~\bibnamefont
  {Akutsu}},\ }\href {https://doi.org/10.1143/JPSJ.56.2248} {\bibfield
  {journal} {\bibinfo  {journal} {J. Phys. Soc. Jpn.}\ }\textbf {\bibinfo
  {volume} {56}},\ \bibinfo {pages} {2248} (\bibinfo {year}
  {1987}{\natexlab{a}})} \BibitemShut {NoStop}%
\bibitem [{\citenamefont {Akutsu}\ and\ \citenamefont
  {Akutsu}(1987{\natexlab{b}})}]{Akutsu1987}%
  \BibitemOpen
  \bibfield  {author} {\bibinfo {author} {\bibfnamefont {Y.}~\bibnamefont
  {Akutsu}}\ and\ \bibinfo {author} {\bibfnamefont {N.}~\bibnamefont
  {Akutsu}},\ }\href {https://doi.org/10.1143/JPSJ.56.9} {\bibfield  {journal}
  {\bibinfo  {journal} {J. Phys. Soc. Jpn.}\ }\textbf {\bibinfo {volume}
  {56}},\ \bibinfo {pages} {9} (\bibinfo {year} {1987}{\natexlab{b}})} \BibitemShut {NoStop}%
\bibitem [{\citenamefont {Holzer}\ and\ \citenamefont
  {Wortis}(1989)}]{Holzer1989}%
  \BibitemOpen
  \bibfield  {author} {\bibinfo {author} {\bibfnamefont {M.}~\bibnamefont
  {Holzer}}\ and\ \bibinfo {author} {\bibfnamefont {M.}~\bibnamefont
  {Wortis}},\ }\href {https://doi.org/10.1103/PhysRevB.40.11044} {\bibfield
  {journal} {\bibinfo  {journal} {Phys. Rev. B}\ }\textbf {\bibinfo {volume}
  {40}},\ \bibinfo {pages} {11044} (\bibinfo {year} {1989})}\BibitemShut
  {NoStop}%
\bibitem [{\citenamefont {Zia}(1978)}]{Zia1978}%
  \BibitemOpen
  \bibfield  {author} {\bibinfo {author} {\bibfnamefont {R.~K.~P.}\
  \bibnamefont {Zia}},\ }\href
  {https://doi.org/https://doi.org/10.1016/0375-9601(78)90261-X} {\bibfield
  {journal} {\bibinfo  {journal} {Physics Letters A}\ }\textbf {\bibinfo
  {volume} {64}},\ \bibinfo {pages} {345 } (\bibinfo {year}
  {1978})}\BibitemShut {NoStop}%
\bibitem [{\citenamefont {Holzer}(1990{\natexlab{a}})}]{Holzer1990}%
  \BibitemOpen
  \bibfield  {author} {\bibinfo {author} {\bibfnamefont {M.}~\bibnamefont
  {Holzer}},\ }\href {https://doi.org/10.1103/PhysRevLett.64.653} {\bibfield
  {journal} {\bibinfo  {journal} {Phys. Rev. Lett.}\ }\textbf {\bibinfo
  {volume} {64}},\ \bibinfo {pages} {653} (\bibinfo {year}
  {1990}{\natexlab{a}})}\BibitemShut {NoStop}%
\bibitem [{\citenamefont {Holzer}(1990{\natexlab{b}})}]{Holzer1990a}%
  \BibitemOpen
  \bibfield  {author} {\bibinfo {author} {\bibfnamefont {M.}~\bibnamefont
  {Holzer}},\ }\href {https://doi.org/10.1103/PhysRevB.42.10570} {\bibfield
  {journal} {\bibinfo  {journal} {Phys. Rev. B}\ }\textbf {\bibinfo {volume}
  {42}},\ \bibinfo {pages} {10570} (\bibinfo {year}
  {1990}{\natexlab{b}})}\BibitemShut {NoStop}%
\bibitem [{\citenamefont {Hamermesh}(1989)}]{Hamermesh1989}%
  \BibitemOpen
  \bibfield  {author} {\bibinfo {author} {\bibfnamefont {M.}~\bibnamefont
  {Hamermesh}},\ }\href {https://doi.org/10.1119/1.1941790} {\emph {\bibinfo
  {title} {Group Theory and Its Application to Physical Problems}}}\ (\bibinfo
  {publisher} {Dover},\ \bibinfo {address} {New York},\ \bibinfo {year}
  {1989})\BibitemShut {NoStop}%
\bibitem [{\citenamefont {Fujimoto}\ and\ \citenamefont
  {Otsuka}(2020)}]{Fujimoto2020}%
  \BibitemOpen
  \bibfield  {author} {\bibinfo {author} {\bibfnamefont {M.}~\bibnamefont
  {Fujimoto}}\ and\ \bibinfo {author} {\bibfnamefont {H.}~\bibnamefont
  {Otsuka}},\ }\href {https://doi.org/10.1103/PhysRevE.102.032141} {\bibfield
  {journal} {\bibinfo  {journal} {Phys. Rev. E}\ }\textbf {\bibinfo {volume}
  {102}},\ \bibinfo {pages} {032141} (\bibinfo {year} {2020})}\BibitemShut
  {NoStop}%
\bibitem [{\citenamefont {Temperley}\ \emph {et~al.}(1971)\citenamefont
  {Temperley}, \citenamefont {Lieb},\ and\ \citenamefont
  {Edwards}}]{Temperley1971}%
  \BibitemOpen
  \bibfield  {author} {\bibinfo {author} {\bibfnamefont {H.~N.~V.}\
  \bibnamefont {Temperley}}, \bibinfo {author} {\bibfnamefont {E.~H.}\
  \bibnamefont {Lieb}},\ and\ \bibinfo {author} {\bibfnamefont {S.~F.}\
  \bibnamefont {Edwards}},\ }\href {https://doi.org/10.1098/rspa.1971.0067}
  {\bibfield  {journal} {\bibinfo  {journal} {Proceedings of the Royal Society
  of London. A. Mathematical and Physical Sciences}\ }\textbf {\bibinfo
  {volume} {322}},\ \bibinfo {pages} {251} (\bibinfo {year} {1971})}
  \BibitemShut {NoStop}%
\bibitem [{\citenamefont {Baxter}(1973)}]{Baxter1973}%
  \BibitemOpen
  \bibfield  {author} {\bibinfo {author} {\bibfnamefont {R.~J.}\ \bibnamefont
  {Baxter}},\ }\href {http://stacks.iop.org/0022-3719/6/i=23/a=005} {\bibfield
  {journal} {\bibinfo  {journal} {J. Phys. C: Solid State Phys.}\ }\textbf
  {\bibinfo {volume} {6}},\ \bibinfo {pages} {L445} (\bibinfo {year}
  {1973})}\BibitemShut {NoStop}%
\bibitem [{\citenamefont {Wu}(1982)}]{Wu1982}%
  \BibitemOpen
  \bibfield  {author} {\bibinfo {author} {\bibfnamefont {F.~Y.}\ \bibnamefont
  {Wu}},\ }\href {https://doi.org/10.1103/RevModPhys.54.235} {\bibfield
  {journal} {\bibinfo  {journal} {Rev. Mod. Phys.}\ }\textbf {\bibinfo {volume}
  {54}},\ \bibinfo {pages} {235} (\bibinfo {year} {1982})}\BibitemShut
  {NoStop}%
\bibitem [{\citenamefont {Kl{\"u}mper}\ \emph {et~al.}(1989)\citenamefont
  {Kl{\"u}mper}, \citenamefont {Schadschneider},\ and\ \citenamefont
  {Zittartz}}]{Kluemper1989}%
  \BibitemOpen
  \bibfield  {author} {\bibinfo {author} {\bibfnamefont {A.}~\bibnamefont
  {Kl{\"u}mper}}, \bibinfo {author} {\bibfnamefont {A.}~\bibnamefont
  {Schadschneider}},\ and\ \bibinfo {author} {\bibfnamefont {J.}~\bibnamefont
  {Zittartz}},\ }\href {https://doi.org/10.1007/BF01312692} {\bibfield
  {journal} {\bibinfo  {journal} {Z. Phys. B}\ }\textbf {\bibinfo {volume}
  {76}},\ \bibinfo {pages} {247} (\bibinfo {year} {1989})}\BibitemShut
  {NoStop}%
\bibitem [{\citenamefont {Buffenoir}\ and\ \citenamefont
  {Wallon}(1993)}]{Buffenoir1993}%
  \BibitemOpen
  \bibfield  {author} {\bibinfo {author} {\bibfnamefont {E.}~\bibnamefont
  {Buffenoir}}\ and\ \bibinfo {author} {\bibfnamefont {S.}~\bibnamefont
  {Wallon}},\ }\href {https://doi.org/10.1088/0305-4470/26/13/009} {\bibfield
  {journal} {\bibinfo  {journal} {J. Phys. A: Math. Gen.}\ }\textbf {\bibinfo
  {volume} {26}},\ \bibinfo {pages} {3045} (\bibinfo {year}
  {1993})}\BibitemShut {NoStop}%
\bibitem [{\citenamefont {Fujimoto}(1990{\natexlab{a}})}]{Fujimoto1990}%
  \BibitemOpen
  \bibfield  {author} {\bibinfo {author} {\bibfnamefont {M.}~\bibnamefont
  {Fujimoto}},\ }\href {https://doi.org/10.1007/BF01334755} {\bibfield
  {journal} {\bibinfo  {journal} {J. Stat. Phys.}\ }\textbf {\bibinfo {volume}
  {59}},\ \bibinfo {pages} {1355} (\bibinfo {year}
  {1990}{\natexlab{a}})}\BibitemShut {NoStop}%
\bibitem [{\citenamefont {Fujimoto}(1990{\natexlab{b}})}]{Fujimoto1990a}%
  \BibitemOpen
  \bibfield  {author} {\bibinfo {author} {\bibfnamefont {M.}~\bibnamefont
  {Fujimoto}},\ }\href@noop {} {\bibfield  {journal} {\bibinfo  {journal} {J.
  Stat. Phys.}\ }\textbf {\bibinfo {volume} {61}},\ \bibinfo {pages} {1295}
  (\bibinfo {year} {1990}{\natexlab{b}})}\BibitemShut {NoStop}%
\bibitem [{\citenamefont {Cheng}\ and\ \citenamefont {Wu}(1967)}]{Cheng1967}%
  \BibitemOpen
  \bibfield  {author} {\bibinfo {author} {\bibfnamefont {H.}~\bibnamefont
  {Cheng}}\ and\ \bibinfo {author} {\bibfnamefont {T.~T.}\ \bibnamefont {Wu}},\
  }\href {https://doi.org/10.1103/PhysRev.164.719} {\bibfield  {journal}
  {\bibinfo  {journal} {Phys. Rev.}\ }\textbf {\bibinfo {volume} {164}},\
  \bibinfo {pages} {719} (\bibinfo {year} {1967})}\BibitemShut {NoStop}%
\bibitem [{\citenamefont {McCoy}\ and\ \citenamefont {Wu}(2013)}]{McCoy2013}%
  \BibitemOpen
  \bibfield  {author} {\bibinfo {author} {\bibfnamefont {B.~M.}\ \bibnamefont
  {McCoy}}\ and\ \bibinfo {author} {\bibfnamefont {T.~T.}\ \bibnamefont {Wu}},\
  }\href {https://doi.org/doi:10.4159/harvard.9780674180758} {\emph {\bibinfo
  {title} {The Two-Dimensional Ising Model}}}\ (\bibinfo  {publisher} {Harvard
  University Press},\ \bibinfo {year} {2013})\BibitemShut {NoStop}%
\bibitem [{\citenamefont {Yamada}(1983)}]{Yamada1983}%
  \BibitemOpen
  \bibfield  {author} {\bibinfo {author} {\bibfnamefont {K.}~\bibnamefont
  {Yamada}},\ }\href {https://doi.org/10.1143/PTP.69.1295} {\bibfield
  {journal} {\bibinfo  {journal} {Prog. Theor. Phys.}\ }\textbf {\bibinfo
  {volume} {69}},\ \bibinfo {pages} {1295} (\bibinfo {year} {1983})}
  \BibitemShut {NoStop}%
\bibitem [{\citenamefont {Yamada}(1984{\natexlab{a}})}]{Yamada1984}%
  \BibitemOpen
  \bibfield  {author} {\bibinfo {author} {\bibfnamefont {K.}~\bibnamefont
  {Yamada}},\ }\href {https://doi.org/10.1143/PTP.71.1416} {\bibfield
  {journal} {\bibinfo  {journal} {Prog. Theor. Phys.}\ }\textbf {\bibinfo
  {volume} {71}},\ \bibinfo {pages} {1416} (\bibinfo {year}
  {1984}{\natexlab{a}})}
  \BibitemShut {NoStop}%
\bibitem [{\citenamefont {Yamada}(1984{\natexlab{b}})}]{Yamada1984a}%
  \BibitemOpen
  \bibfield  {author} {\bibinfo {author} {\bibfnamefont {K.}~\bibnamefont
  {Yamada}},\ }\href {https://doi.org/10.1143/PTP.72.922} {\bibfield  {journal}
  {\bibinfo  {journal} {Prog. Theor. Phys.}\ }\textbf {\bibinfo {volume}
  {72}},\ \bibinfo {pages} {922} (\bibinfo {year} {1984}{\natexlab{b}})}
  \BibitemShut {NoStop}%
\bibitem [{\citenamefont {Yamada}(1986)}]{Yamada1986}%
  \BibitemOpen
  \bibfield  {author} {\bibinfo {author} {\bibfnamefont {K.}~\bibnamefont
  {Yamada}},\ }\href {https://doi.org/10.1143/PTP.76.602} {\bibfield  {journal}
  {\bibinfo  {journal} {Prog. Theor. Phys.}\ }\textbf {\bibinfo {volume}
  {76}},\ \bibinfo {pages} {602} (\bibinfo {year} {1986})}
  \BibitemShut {NoStop}%
\bibitem [{\citenamefont {Johnson}\ \emph {et~al.}(1973)\citenamefont
  {Johnson}, \citenamefont {Krinsky},\ and\ \citenamefont
  {McCoy}}]{Johnson1973}%
  \BibitemOpen
  \bibfield  {author} {\bibinfo {author} {\bibfnamefont {J.~D.}\ \bibnamefont
  {Johnson}}, \bibinfo {author} {\bibfnamefont {S.}~\bibnamefont {Krinsky}},\
  and\ \bibinfo {author} {\bibfnamefont {B.~M.}\ \bibnamefont {McCoy}},\ }\href
  {https://doi.org/10.1103/PhysRevA.8.2526} {\bibfield  {journal} {\bibinfo
  {journal} {Phys. Rev. A}\ }\textbf {\bibinfo {volume} {8}},\ \bibinfo {pages}
  {2526} (\bibinfo {year} {1973})}\BibitemShut {NoStop}%
\bibitem [{\citenamefont {Laanait}(1987)}]{Laanait1987}%
  \BibitemOpen
  \bibfield  {author} {\bibinfo {author} {\bibfnamefont {L.}~\bibnamefont
  {Laanait}},\ }\href
  {https://doi.org/https://doi.org/10.1016/0375-9601(87)90048-X} {\bibfield
  {journal} {\bibinfo  {journal} {Physics Letters A}\ }\textbf {\bibinfo
  {volume} {124}},\ \bibinfo {pages} {480 } (\bibinfo {year}
  {1987})}\BibitemShut {NoStop}%
\bibitem [{\citenamefont {Fujimoto}(1996)}]{Fujimoto1996}%
  \BibitemOpen
  \bibfield  {author} {\bibinfo {author} {\bibfnamefont {M.}~\bibnamefont
  {Fujimoto}},\ }\href
  {https://doi.org/https://doi.org/10.1016/S0378-4371(96)00224-5} {\bibfield
  {journal} {\bibinfo  {journal} {Physica A}\ }\textbf {\bibinfo {volume}
  {233}},\ \bibinfo {pages} {485 } (\bibinfo {year} {1996})}\BibitemShut
  {NoStop}%
\bibitem [{\citenamefont {Vaidya}(1976)}]{Vaidya1976}%
  \BibitemOpen
  \bibfield  {author} {\bibinfo {author} {\bibfnamefont {H.~G.}\ \bibnamefont
  {Vaidya}},\ }\href
  {https://doi.org/https://doi.org/10.1016/0375-9601(76)90432-1} {\bibfield
  {journal} {\bibinfo  {journal} {Physics Letters A}\ }\textbf {\bibinfo
  {volume} {57}},\ \bibinfo {pages} {1 } (\bibinfo {year} {1976})}\BibitemShut
  {NoStop}%
\bibitem [{\citenamefont {Stephenson}(1964)}]{Stephenson1964}%
  \BibitemOpen
  \bibfield  {author} {\bibinfo {author} {\bibfnamefont {J.}~\bibnamefont
  {Stephenson}},\ }\href@noop {} {\bibfield  {journal} {\bibinfo  {journal}
  {Journal of Mathematical Physics}\ }\textbf {\bibinfo {volume} {5}},\
  \bibinfo {pages} {1009} (\bibinfo {year} {1964})}\BibitemShut {NoStop}%
\bibitem [{\citenamefont {Chan}\ \emph {et~al.}(2011)\citenamefont {Chan},
  \citenamefont {Guttmann}, \citenamefont {Nickel},\ and\ \citenamefont
  {Perk}}]{Chan2011}%
  \BibitemOpen
  \bibfield  {author} {\bibinfo {author} {\bibfnamefont {Y.}~\bibnamefont
  {Chan}}, \bibinfo {author} {\bibfnamefont {A.~J.}\ \bibnamefont {Guttmann}},
  \bibinfo {author} {\bibfnamefont {B.~G.}\ \bibnamefont {Nickel}},\ and\
  \bibinfo {author} {\bibfnamefont {J.~H.~H.}\ \bibnamefont {Perk}},\ }\href
  {https://doi.org/10.1007/s10955-011-0212-0} {\bibfield  {journal} {\bibinfo
  {journal} {J. Stat. Phys.}\ }\textbf {\bibinfo {volume} {145}},\ \bibinfo
  {pages} {549} (\bibinfo {year} {2011})},\ \bibinfo {note} {and the references
  therein}\BibitemShut {NoStop}%
\bibitem [{\citenamefont {Fujimoto}(1999)}]{Fujimoto1999}%
  \BibitemOpen
  \bibfield  {author} {\bibinfo {author} {\bibfnamefont {M.}~\bibnamefont
  {Fujimoto}},\ }\href
  {https://doi.org/https://doi.org/10.1016/S0378-4371(98)00393-8} {\bibfield
  {journal} {\bibinfo  {journal} {Physica A}\ }\textbf {\bibinfo {volume}
  {264}},\ \bibinfo {pages} {149 } (\bibinfo {year} {1999})}\BibitemShut
  {NoStop}%
\bibitem [{\citenamefont {Fujimoto}(2002{\natexlab{a}})}]{Fujimoto2002}%
  \BibitemOpen
  \bibfield  {author} {\bibinfo {author} {\bibfnamefont {M.}~\bibnamefont
  {Fujimoto}},\ }\href {http://stacks.iop.org/0305-4470/35/i=34/a=402}
  {\bibfield  {journal} {\bibinfo  {journal} {J. Phys. A: Math. Gen.}\ }\textbf
  {\bibinfo {volume} {35}},\ \bibinfo {pages} {7553} (\bibinfo {year}
  {2002}{\natexlab{a}})}\BibitemShut {NoStop}%
\bibitem [{\citenamefont {{Walker}}(1950)}]{Walker1950}%
  \BibitemOpen
  \bibfield  {author} {\bibinfo {author} {\bibfnamefont {R.~J.}\ \bibnamefont
  {{Walker}}},\ }\href@noop {} {\emph {\bibinfo {title} {Algebraic curves}}},\
  Vol.~\bibinfo {volume} {13}\ (\bibinfo  {publisher} {Princeton University
  Press, Princeton, NJ},\ \bibinfo {year} {1950})\BibitemShut {NoStop}%
\bibitem [{\citenamefont {Zia}(1986)}]{Zia1986}%
  \BibitemOpen
  \bibfield  {author} {\bibinfo {author} {\bibfnamefont {R.~K.~P.}\
  \bibnamefont {Zia}},\ }\href {https://doi.org/10.1007/BF01020575} {\bibfield
  {journal} {\bibinfo  {journal} {J. Stat. Phys.}\ }\textbf {\bibinfo {volume}
  {45}},\ \bibinfo {pages} {801} (\bibinfo {year} {1986})}\BibitemShut
  {NoStop}%
\bibitem [{\citenamefont {Swendsen}\ and\ \citenamefont
  {Wang}(1987)}]{Swendsen1987}%
  \BibitemOpen
  \bibfield  {author} {\bibinfo {author} {\bibfnamefont {R.~H.}\ \bibnamefont
  {Swendsen}}\ and\ \bibinfo {author} {\bibfnamefont {J.-S.}\ \bibnamefont
  {Wang}},\ }\href {https://doi.org/10.1103/PhysRevLett.58.86} {\bibfield
  {journal} {\bibinfo  {journal} {Phys. Rev. Lett.}\ }\textbf {\bibinfo
  {volume} {58}},\ \bibinfo {pages} {86} (\bibinfo {year} {1987})}\BibitemShut
  {NoStop}%
\bibitem [{\citenamefont {Wolff}(1988)}]{Wolff1988}%
  \BibitemOpen
  \bibfield  {author} {\bibinfo {author} {\bibfnamefont {U.}~\bibnamefont
  {Wolff}},\ }\href {https://doi.org/10.1103/PhysRevLett.60.1461} {\bibfield
  {journal} {\bibinfo  {journal} {Phys. Rev. Lett.}\ }\textbf {\bibinfo
  {volume} {60}},\ \bibinfo {pages} {1461} (\bibinfo {year}
  {1988})}\BibitemShut {NoStop}%
\bibitem [{\citenamefont {Wolff}(1989)}]{Wolff1989}%
  \BibitemOpen
  \bibfield  {author} {\bibinfo {author} {\bibfnamefont {U.}~\bibnamefont
  {Wolff}},\ }\href {https://doi.org/10.1103/PhysRevLett.62.361} {\bibfield
  {journal} {\bibinfo  {journal} {Phys. Rev. Lett.}\ }\textbf {\bibinfo
  {volume} {62}},\ \bibinfo {pages} {361} (\bibinfo {year} {1989})}\BibitemShut
  {NoStop}%
\bibitem [{\citenamefont {Evertz}\ and\ \citenamefont {von~der
  Linden}(2001)}]{Evertz2001}%
  \BibitemOpen
  \bibfield  {author} {\bibinfo {author} {\bibfnamefont {H.~G.}\ \bibnamefont
  {Evertz}}\ and\ \bibinfo {author} {\bibfnamefont {W.}~\bibnamefont {von~der
  Linden}},\ }\href {https://doi.org/10.1103/PhysRevLett.86.5164} {\bibfield
  {journal} {\bibinfo  {journal} {Phys. Rev. Lett.}\ }\textbf {\bibinfo
  {volume} {86}},\ \bibinfo {pages} {5164} (\bibinfo {year}
  {2001})}\BibitemShut {NoStop}%
\bibitem [{\citenamefont {Fortuin}\ and\ \citenamefont
  {Kasteleyn}(1972)}]{Fortuin1972}%
  \BibitemOpen
  \bibfield  {author} {\bibinfo {author} {\bibfnamefont {C.~M.}\ \bibnamefont
  {Fortuin}}\ and\ \bibinfo {author} {\bibfnamefont {P.~W.}\ \bibnamefont
  {Kasteleyn}},\ }\href {https://doi.org/10.1016/0031-8914(72)90045-6}
  {\bibfield  {journal} {\bibinfo  {journal} {Physica}\ }\textbf {\bibinfo
  {volume} {57}},\ \bibinfo {pages} {536} (\bibinfo {year} {1972})}\BibitemShut
  {NoStop}%
\bibitem [{\citenamefont {Suzuki}(1974)}]{Suzuki1974}%
  \BibitemOpen
  \bibfield  {author} {\bibinfo {author} {\bibfnamefont {M.}~\bibnamefont
  {Suzuki}},\ }\href {https://doi.org/10.1143/PTP.51.1992} {\bibfield
  {journal} {\bibinfo  {journal} {Prog. Theor. Phys.}\ }\textbf {\bibinfo
  {volume} {51}},\ \bibinfo {pages} {1992} (\bibinfo {year}
  {1974})}\BibitemShut {NoStop}%
\bibitem [{\citenamefont {Fujimoto}(2002{\natexlab{b}})}]{Fujimoto2002a}%
  \BibitemOpen
  \bibfield  {author} {\bibinfo {author} {\bibfnamefont {M.}~\bibnamefont
  {Fujimoto}},\ }\href {http://stacks.iop.org/0305-4470/35/i=7/a=304}
  {\bibfield  {journal} {\bibinfo  {journal} {J. Phys. A: Math. Gen.}\ }\textbf
  {\bibinfo {volume} {35}},\ \bibinfo {pages} {1517} (\bibinfo {year}
  {2002}{\natexlab{b}})}\BibitemShut {NoStop}%
\bibitem [{\citenamefont {Namba}(1984)}]{Namba1984}%
  \BibitemOpen
  \bibfield  {author} {\bibinfo {author} {\bibfnamefont {M.}~\bibnamefont
  {Namba}},\ }\href {https://books.google.co.jp/books?id=IkXvAAAAMAAJ} {\emph
  {\bibinfo {title} {Geometry of Projective Algebraic Curves}}},\ Monographs
  and textbooks in pure and applied mathematics\ (\bibinfo  {publisher} {M.
  Dekker},\ \bibinfo {address} {New York},\ \bibinfo {year} {1984})\BibitemShut
  {NoStop}%
\bibitem [{\citenamefont {Baxter}(1978)}]{Baxter1978}%
  \BibitemOpen
  \bibfield  {author} {\bibinfo {author} {\bibfnamefont {R.~J.}\ \bibnamefont
  {Baxter}},\ }\href {https://doi.org/10.1098/rsta.1978.0062} {\bibfield
  {journal} {\bibinfo  {journal} {Philos. Trans. R. Soc. Lond., Ser. A}\
  }\textbf {\bibinfo {volume} {289}},\ \bibinfo {pages} {315} (\bibinfo {year}
  {1978})}\BibitemShut {NoStop}%
\bibitem [{\citenamefont {Fujimoto}\ and\ \citenamefont
  {Otsuka}(2022)}]{Fujimoto2020e}%
  \BibitemOpen
  \bibfield  {author} {\bibinfo {author} {\bibfnamefont {M.}~\bibnamefont
  {Fujimoto}}\ and\ \bibinfo {author} {\bibfnamefont {H.}~\bibnamefont
  {Otsuka}},\ }\href {https://doi.org/10.1103/PhysRevE.105.059904} {\bibfield
  {journal} {\bibinfo  {journal} {Phys. Rev. E}\ }\textbf {\bibinfo {volume}
  {105}},\ \bibinfo {pages} {059904} (\bibinfo {year} {2022})}\BibitemShut
  {NoStop}%
\end{thebibliography}
%

 \end{document}